\documentclass[letterpaper]{article} % DO NOT CHANGE THIS
\usepackage[draft]{aaai2026}  % DO NOT CHANGE THIS
\usepackage{times}  % DO NOT CHANGE THIS
\usepackage{helvet}  % DO NOT CHANGE THIS
\usepackage{courier}  % DO NOT CHANGE THIS
\usepackage[hyphens]{url}  % DO NOT CHANGE THIS
\usepackage{graphicx} % DO NOT CHANGE THIS
\usepackage{natbib}  % DO NOT CHANGE THIS AND DO NOT ADD ANY OPTIONS TO IT
\usepackage{caption} % DO NOT CHANGE THIS AND DO NOT ADD ANY OPTIONS TO IT
\usepackage{algorithm}
\usepackage{newfloat}
\usepackage{listings}
\usepackage{pifont}
\usepackage{amsmath}
\usepackage{amssymb}
\usepackage{algpseudocode}
\usepackage{booktabs}
\usepackage{multirow}
\definecolor{darkgreen}{rgb}{0.0, 0.5, 0.0}

\newtheorem{lemma}{Lemma}

\DeclareCaptionStyle{ruled}{labelfont=normalfont,labelsep=colon,strut=off} % DO NOT CHANGE THIS
\floatstyle{ruled}
\newfloat{listing}{tb}{lst}{}
\floatname{listing}{Listing}
\title{SelectiveShield: Lightweight Hybrid Defense Against Gradient Leakage in Federated Learning}
\author{
   Borui Li,
   Li Yan,
   Jianmin Liu
}
\affiliations{
     School of Cyber Science and Engineering\\
      Xi'an Jiaotong University\\
      Shaanxi, China 710049 \\

    boruili@stu.xjtu.edu.cn
}
\usepackage{bibentry}

\begin{document}

\maketitle

\begin{abstract}
Federated Learning (FL) enables collaborative model training on decentralized data but remains vulnerable to gradient leakage attacks that can reconstruct sensitive user information. Existing defense mechanisms, such as differential privacy (DP) and homomorphic encryption (HE), often introduce a trade-off between privacy, model utility, and system overhead, a challenge that is exacerbated in heterogeneous environments with non-IID data and varying client capabilities. To address these limitations, we propose SelectiveShield, a lightweight hybrid defense framework that adaptively integrates selective homomorphic encryption and differential privacy. SelectiveShield leverages Fisher information to quantify parameter sensitivity, allowing clients to identify critical parameters locally. Through a collaborative negotiation protocol, clients agree on a shared set of the most sensitive parameters for protection via homomorphic encryption. Parameters that are uniquely important to individual clients are retained locally, fostering personalization, while non-critical parameters are protected with adaptive differential privacy noise. Extensive experiments demonstrate that SelectiveShield maintains strong model utility while significantly mitigating gradient leakage risks, offering a practical and scalable defense mechanism for real-world federated learning deployments.
\end{abstract}

\section{Introduction}
Federated Learning FL has emerged as a decentralized machine learning paradigm that enables collaborative model training while preserving data locality~\cite{mcmahan2017communication,kairouz2021advances}. Despite its privacy-centric design, FL remains vulnerable to inversion attacks that exploit shared gradients or model updates to reconstruct sensitive information ~\cite{Huang2021Evaluati,guo2025exploring}. These attacks threaten data confidentiality by enabling adversaries to infer private attributes or reconstruct raw training samples without direct data access, undermining FL's core privacy guarantees.

Current defense mechanisms against inversion attacks in FL predominantly employ Differential Privacy(DP), Homomorphic Encryption(HE), and Secure Multi-Party Computation (SMPC). 
Differential privacy injects calibrated noise into gradients to protect sensitive information~\cite{shokri2015privacy}, but this often introduces a trade-off between privacy and model utility, as excessive noise can significantly impair performance ~\cite{Wang2024More}. In addition, DP-based methods necessitate a substantial participant pool during training to achieve convergence and attain an optimal privacy-performance balance ~\cite{sun2021soteria}.
SMPC enables secure computation over encrypted data, allowing participants to jointly compute model updates without revealing individual gradients~\cite{bonawitz2017practical,wu2024federated}. However, it incurs substantial communication and computational overhead, making it less practical for large-scale FL deployments.
Moreover, SMPC imposes stringent requirements on the number of participating clients per round and demonstrates limited robustness in handling client dropout scenarios.
Homomorphic Encryption safeguards privacy by encrypting model updates before transmission~\cite{kumar2024revamping,zhou2024two}. However, using full homomorphic encryption for all model updates leads to significant computational and communication overhead~\cite{yan2024efficient}, which limits its practicality in large-scale or resource-constrained federated learning scenarios.

Primary privacy methods present a sharp trade-off. Homomorphic Encryption preserves model accuracy but incurs significant computational overhead, whereas Differential Privacy is highly efficient but can degrade model performance. This prompts the question: \textbf{\ding{182}  How can a hybrid system be designed to optimally allocate the application of these defenses. } Specifically, can we use a parameter importance metric like Fisher Information to apply accuracy-preserving Homomorphic Encryption exclusively to the most critical parameters, while protecting the numerous less-sensitive parameters with efficient Differential Privacy noise, where their impact on model performance is minimal?

In statistically heterogeneous federated learning scenarios, clients typically exhibit non-independent and identically distributed (non-IID) data characteristics. These clients often have significantly varying datasets in terms of volume and class distribution. This leads to divergence in local models, where the most sensitive parameters critical to model performance vary across different clients. Consequently, each client may generate a distinct local mask to protect its unique set of important parameters.

\begin{figure}
    \centering
    \includegraphics[width=0.5\linewidth]{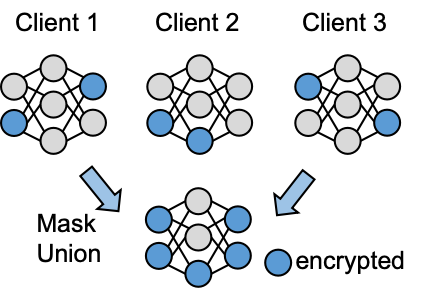}
    \caption{The Challenge of Selective HE Method}
    \label{fig:Challenge}
\end{figure}

As depicted in Figure \ref{fig:Challenge}, clients in a federated learning system identify different sets of sensitive parameters (highlighted in blue) based on their local, heterogeneous data. This divergence creates a fundamental conflict with homomorphic encryption, which requires all participants to use a common encryption mask for correct aggregation and decryption. A naive solution, such as taking the union of all local masks, is therefore untenable. In highly heterogeneous settings, this would cause the global encryption mask to cover a significantly larger portion of the model, dramatically increasing computational overhead and negating the benefits of a selective approach. This brings us to our second, highly-focused question:
\textbf{\ding{183} How can a selective defense protocol resolve the operational conflicts arising from heterogeneity?}

To address this issue, we draw inspiration from personalized federated learning by proposing a parameter partitioning strategy. After each client determines its local sensitivity mask, all participants negotiate to establish a common encryption mask defined as the intersection of their individual masks. This alignment on a shared set of parameters is crucial, as it enables the use of efficient packing-based homomorphic encryption schemes, accelerating the cryptographic process.

Parameters that are sensitive to individual clients but fall outside this common intersection are designated as personalized knowledge. These parameters are retained on the client side and are excluded from the current round of global aggregation. This approach of retaining client-specific parameters locally yields significant benefits: it reduces communication and computational overhead by shrinking the encrypted set, mitigates non-IID issues and preserves high model utility even in settings with significant data heterogeneity. As our experimental results confirm, this strategy allows our framework to achieve accuracy that is comparable or even superior to conventional methods.

Furthermore, many existing FL+HE schemes rely on a shared private key distributed among all clients, introducing a critical vulnerability: a single malicious participant could decrypt any other client's updates, completely nullifying the privacy guarantee. This raises our third key question: \textbf{\ding{184} How can we design a secure aggregation protocol for selective homomorphic encryption where no private key is ever exposed to the clients, thus protecting encrypted updates even from malicious insider attacks?
}

To address these challenges, we propose \textbf{SelectiveShield}, a lightweight hybrid defense framework that provides a multi-faceted solution. SelectiveShield uses Fisher Information to systematically partition model parameters based on their sensitivity. Answering our first question, it applies accuracy-preserving homomorphic encryption only to a small, collaboratively-agreed set of critical parameters, while protecting the vast majority of less impactful parameters with efficient differential privacy noise. To resolve the conflicts from heterogeneity, it uses a negotiation protocol to establish a common encryption domain, while treating client-specific sensitive parameters as personalized knowledge that is retained locally, thus boosting performance in non-IID settings. Finally, to prevent insider attacks, it employs a two-server architecture where the private key is exclusively managed by a trusted server and is never exposed to clients, ensuring end-to-end confidentiality.

\section{Related Work}

\subsection{Privacy-Preserving Techniques in Federated Learning}
Despite keeping data localized, FL remains vulnerable to various privacy attacks, including gradient inversion attacks that can reconstruct raw training data~\cite{zhu2019deep, geiping2020inverting}, membership inference attacks that determine whether specific samples were used in training~\cite{melis2019exploiting, shokri2017membership}, and property inference attacks that extract sensitive statistical or semantic attributes~\cite{wang2022poisoning, luo2021feature}.

Differential privacy (DP) in FL typically involves adding calibrated noise to model updates before sharing them with the server \cite{geyer2017differentially, wei2020federated}. Truex et al. \cite{truex2020ldp} proposed a hybrid approach combining local and central differential privacy. Yang et al. \cite{yang2023dynamic} introduced dynamic personalized FL with adaptive differential privacy, using Fisher information to guide the noise addition process. While DP provides theoretical privacy guarantees, it often degrades model performance, especially with stronger privacy requirements.

Homomorphic encryption (HE) enables computations on encrypted data without decryption \cite{phong2018privacy, zhang2020batchcrypt}. Phong et al. \cite{phong2018privacy} applied additive HE to protect client updates in FL. Zhang et al. \cite{zhang2020batchcrypt} proposed BatchCrypt to reduce the overhead of HE through quantization and batching. 
Jin et al. proposed FedML-HE, a homomorphic encryption-based federated learning system that uses selective parameter encryption to reduce computational and communication overheads while preserving privacy.
Hu and Li \cite{hu2024maskcrypt} introduced MaskCrypt, which selectively encrypts only a portion of model updates using gradient-guided masks, significantly reducing communication overhead while maintaining protection against privacy attacks. 
\section{Preliminary}

\subsection{Threat Model}
We consider an \textit{honest-but-curious} server that strictly adheres to the federated learning aggregation protocol while attempting to infer sensitive information from the received gradients or model updates. Although clients faithfully compute their local updates, they remain vulnerable to privacy breaches through the shared parameters. Malicious adversaries may attempt to reconstruct raw training data from the model updates through gradient inversion attacks, formulated as solving the following optimization problem:

\begin{equation}
    \min_{\hat{\mathbf{x}}} \; \Big\| \nabla_{\theta} \mathcal{L}(\theta, \hat{\mathbf{x}}, y) - \mathbf{g} \Big\|^2
\end{equation}

where $\hat{\mathbf{x}}$ represents the reconstructed input data, $\theta$ denotes the model parameters, $\mathcal{L}$ is the loss function, and $\mathbf{g}$ is the observed gradient.

\subsection{Fisher Information}
Fisher Information (FI) measures how much information an observable variable carries about an unknown parameter. For a parametric model with likelihood function $p(x|\theta)$, the Fisher Information Matrix (FIM) is:
\begin{equation}
    \mathbf{I}(\theta) = \mathbb{E}_{x \sim p(x|\theta)}
    \Big[ (\nabla_{\theta} \log p(x|\theta))
    (\nabla_{\theta} \log p(x|\theta))^{\top} \Big].
\end{equation}
In practice, a diagonal approximation is often used:
\begin{equation}
    I_{j}(\theta) \approx \frac{1}{N} \sum_{i=1}^{N} 
    \Bigg( \frac{\partial}{\partial \theta_j} 
    \log \mathcal{L}(x_i, y_i; \theta) \Bigg)^2.
\end{equation}
Parameters with higher FI are more sensitive and should be protected more rigorously.

\subsection{R\'enyi Differential Privacy}
R\'enyi Differential Privacy (RDP)~\cite{mironov2017renyi} extends classical DP by using R\'enyi divergence to obtain tighter composition bounds. For two probability distributions $P$ and $Q$, their R\'enyi divergence of order $\alpha > 1$ is defined as:
\begin{equation}
    D_{\alpha}(P \| Q) = 
    \frac{1}{\alpha - 1} 
    \log \mathbb{E}_{x \sim Q} 
    \Bigg[ \Big( \frac{P(x)}{Q(x)} \Big)^{\alpha} \Bigg].
\end{equation}
A randomized mechanism $\mathcal{M}$ satisfies $(\alpha, \epsilon)$-RDP if for all neighboring datasets $D$ and $D'$:
\begin{equation}
    D_{\alpha}(\mathcal{M}(D) \| \mathcal{M}(D')) \leq \epsilon.
\end{equation}
The Gaussian mechanism achieves RDP by adding noise:
\begin{equation}
    \mathcal{M}(x) = f(x) + \mathcal{N}(0, \sigma^2 S^2),
\end{equation}
where $S$ is the global sensitivity and $\sigma$ controls the privacy budget.

\begin{lemma}[RDP composition \cite{mironov2017renyi}]
  \label{lemma:rdp_composition}
  If $\mathcal{M}_{1}$ satisfies $(\alpha, \epsilon_1)$-RDP and $\mathcal{M}_{2}$ satisfies $(\alpha, \epsilon_2)$, then their
  composition $\mathcal{M}_{1} \circ \mathcal{M}_{2}$ satisfies $(\alpha, \epsilon_1 + \epsilon_2)$-RDP.
\end{lemma}
  
\begin{lemma}[RDP to DP conversion \cite{balle2020hypothesis}]
  \label{lemma:rdp_conversion}
  If the mechanism $\mathcal{M}$ satisfies $(\alpha, \epsilon)$-Rényi differential privacy (RDP), then it also satisfies $(\epsilon', \delta)$-differential privacy (DP) for any $0 < \delta < 1$ where
  {\small
  \begin{equation}
    \nonumber
    \epsilon' = \epsilon + \log{\frac{\alpha-1}{\alpha}} - \frac{\log{\delta} + \log{\alpha}}{\alpha-1}.
  \end{equation}
  }
\end{lemma}

\begin{lemma}[RDP Gaussian mechanism \cite{mironov2017renyi}]  
\label{lemma:rdp_gaussian}
If a function $f: D \rightarrow \mathbb{R}^d$ has $\ell_2$-sensitivity $\Delta_{f}$, then the Gaussian mechanism $G_{f}(\cdot) := f(\cdot) + \mathcal{N}(0, \sigma^2\Delta_{f}^2 I)$ satisfies $(\alpha, \alpha / (2\sigma^2))$-Rényi differential privacy (RDP) for any $\alpha > 1$.
\end{lemma}

\subsection{Homomorphic Encryption}
Homomorphic Encryption (HE) allows computations to be performed directly on encrypted data.  
A typical additively homomorphic encryption scheme (e.g., Paillier) supports operations like:
\begin{equation}
    Enc(m_1) + Enc(m_2) = Enc(m_1 + m_2).
\end{equation}
This property enables secure aggregation in FL:
\begin{equation}
    Enc\Big( \sum_{i=1}^{M} w_i \Big) = \sum_{i=1}^{M} Enc(w_i),
\end{equation}
allowing the server to sum encrypted updates without decrypting individual client gradients.

In contrast, Fully Homomorphic Encryption (FHE) schemes, such as BGV or CKKS, support both addition and multiplication:
\begin{equation}
    Enc(m_1) \cdot Enc(m_2) = Enc(m_1 \cdot m_2).
\end{equation}
CKKS is widely used in machine learning because it efficiently handles approximate real-number arithmetic, which is common for gradient updates.  
However, FHE schemes come with significant computational and communication overheads, making them impractical for encrypting all model parameters in large-scale FL. Therefore, selective or partial encryption strategies are commonly adopted to balance privacy and efficiency.

\subsection{Personalized Federated Learning}
Personalized Federated Learning (PFL) addresses the non-IID nature of clients’ local data by separating model parameters into a global component $\mathbf{v}$ and a client-specific local component $\mathbf{u}_i$. Suppose client $i$ holds $D_i = \{(\mathbf{x}_j, y_j)\}_{j=1}^{N_i}$. The objective function is:
\begin{equation}
    \min_{\mathbf{v}, \mathbf{u}_{1:M}} 
    \frac{1}{M} \sum_{i=1}^{M} 
    \mathbb{E}_{(\mathbf{x}, y) \sim D_i} 
    \Big[ \mathcal{L}(f(\mathbf{x}; \mathbf{v}, \mathbf{u}_i), y) \Big].
\end{equation}
Each client locally optimizes its parameters and then only shares updates related to the global component:
\begin{equation}
    \mathbf{v}^{t+1} = \mathbf{v}^{t} + 
    \frac{1}{M} \sum_{i=1}^{M} 
    (\mathbf{v}_i^{t+1} - \mathbf{v}^{t}).
\end{equation}
PFL frameworks support better personalization and generalization on heterogeneous data. In our method, the separation of sensitive and non-sensitive parameters naturally aligns with this idea: parameters outside the shared secure mask can be treated as local knowledge and remain private.

\section{Method}
\begin{figure}
    \centering
    \includegraphics[width=1\linewidth]{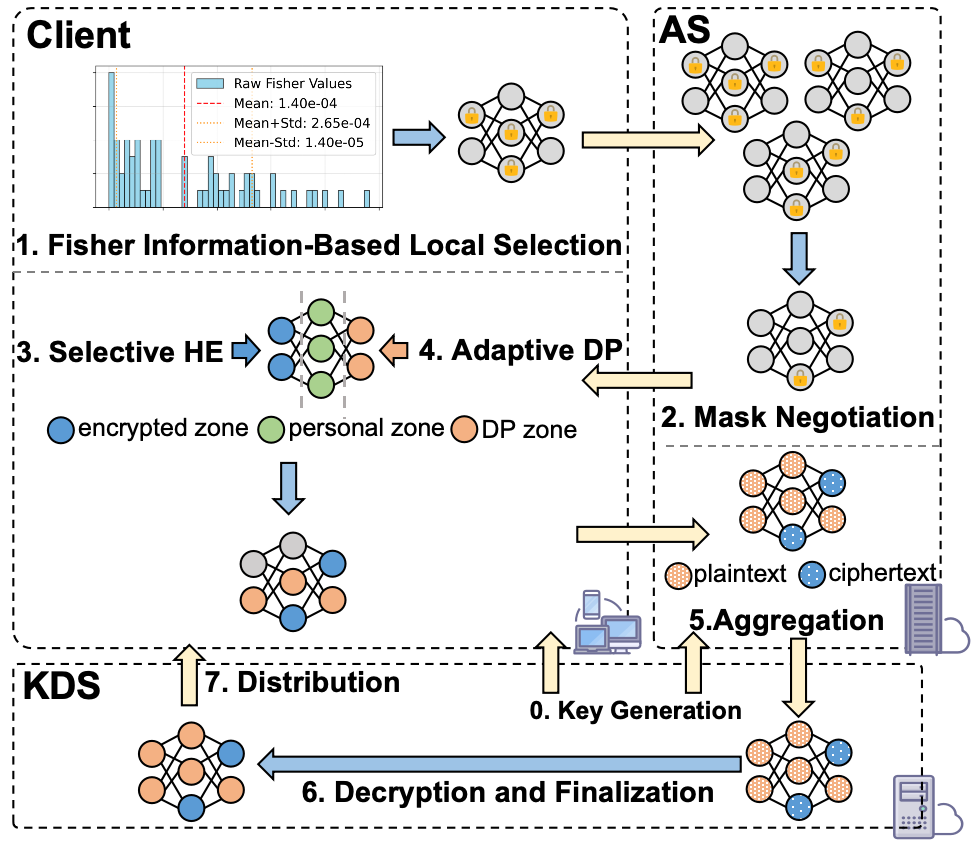}
    \caption{Framework of SelectiveShield}
    \label{fig:method}
\end{figure}

We introduce SelectiveShield, an adaptive hybrid defense framework, with its comprehensive workflow depicted in Figure \ref{fig:method}.
The process begins with each client using Fisher Information to locally identify sensitive parameters (Section \ref{sec:fisher}). Following this, clients engage in a collaborative negotiation protocol (Section \ref{sec:collaborative}) to partition the model's parameters into three zones based on a shared consensus: a globally-critical encrypted zone that is shared, a personalized zone that is retained locally, and a noise zone that is protected via differential privacy. Clients then apply a hybrid defense, protecting the encrypted zone with homomorphic encryption (Section \ref{sec:enc}) and the noise zone with differential privacy (Section \ref{sec:noise}), before uploading the updates. To ensure security, a two-server architecture (Section \ref{sec:agg}) is employed, consisting of a trusted \textbf{Key Distribution and Decryption Server (KDS, $S_\text{Key}$)} responsible for public key distribution and model decryption, along with a semi-trusted \textbf{Aggregation Server (AS, $S_\text{Agg}$)} dedicated to aggregating encrypted models.
The updated global model is then distributed back to the clients, who merge it with their personalized parameters for the next training round.

We consider $T$ communication rounds, each with a set of participating clients $K^t$. Let $N$ be the total number of clients. Clients are selected in each round via Poisson sampling with selection probability $q_n^t$. During the initialization phase, $S_\text{Key}$ distributes the public key $pk$ to all clients and $S_\text{Agg}$.

\begin{algorithm}[t]
\caption{SelectiveShield}
\label{alg:selectiveshield_concise}
\begin{algorithmic}[1]
\State \textbf{Init:} $S_\text{Key}: (pk, sk) \gets \text{KeyGen}(), \theta_g^0 \gets \text{Init}()$;
\State \hspace{0.8cm} $S_\text{Key}$ sends $pk$ to $S_\text{Agg}, $ all clients.
\For{$t = 1, \dots, T$}
    \State $K^t \gets \text{SelectClients()}$

    \Statex
    \State \textbf{Client} $n \in K^t$ ():
    \State \hspace{0.5cm} $\Delta\theta_n^t \gets \text{LocalTrain}(\theta_n^{t-1}, D_n)$
    \State \hspace{0.5cm} $M_n^t \gets \text{GenMask}(\Delta\theta_n^t, \tau_n)$ 
    \Comment{Refer to Sec \ref{sec:fisher}}

    \Statex
    \State \textbf{Mask Negotiation()}:
    \State \hspace{0.5cm} $M_{\text{enc}}^t \gets \bigcap_{n \in K^t} M_n^t$;  
    \Comment{Encrypted}
    \State \hspace{0.5cm} $M_{\text{pers}}^t \gets M_{n}^t \setminus M_{\text{enc}}^t$; \Comment{Personalized}
    \State \hspace{0.5cm} $M_{\text{noise}}^t \gets \Theta \setminus 
    \Big( M_{\text{enc}}^t \cup M_{\text{pers}}^t \Big)$ \Comment{Noised}

    \Statex
    \State \textbf{Client} $n \in K^t$ ():
    \State \hspace{0.5cm} $\mathbf{c}_n^t \gets \text{Enc}(pk, \Delta\theta_{n, M_{\text{enc}}^t})$
    \State \hspace{0.5cm} $\tilde{\Delta\theta}_{n}^t \gets \text{AddNoise}(\Delta\theta_{n, M_{\text{noise}}^t}, C_n^t, \sigma_n^t)$ 

    \Statex
    \State \textbf{Server $S_\text{Agg}()$}:
    \State \hspace{0.5cm}  $\mathbf{c}_{\text{agg}}^t \gets \sum_{n \in K^t} \mathbf{c}_n^t$; 
    \State \hspace{0.5cm} $\Delta\theta_{\text{noise\_sum}}^t \gets \sum_{n \in K^t} \tilde{\Delta\theta}_{n}^t$

    \Statex
    \State \textbf{Server $S_\text{Key}()$}:
    \State \hspace{0.5cm}  $\Delta\theta_{\text{enc\_sum}}^t \gets \text{Dec}(sk, \mathbf{c}_{\text{agg}}^t)$
    \State \hspace{0.5cm}  $\Delta\theta_{\text{sum}}^t \gets \Delta\theta_{\text{enc\_sum}}^t \oplus \Delta\theta_{\text{noise\_sum}}^t$
    \State \hspace{0.5cm} $\theta_\text{global}^t \gets \theta_{\text{global}}^{t-1} + \eta_g \frac{1}{|K^t|} \Delta\theta_{\text{sum}}^t$

    \Statex
    \State \textbf{Client} $n \in K^t$ ():
    \State \hspace{0.5cm}  $\theta_n^{t} \gets {M_{\text{pers}}^t} \odot \theta_{n,\text{local}}^{t} + (1 - {M_{\text{pers}}^t}) \odot \theta_\text{global}^t$
\EndFor
\end{algorithmic}
\end{algorithm}

\subsection{Fisher Information-Based Local Selection}
\label{sec:fisher}

At the beginning of each round $t$, each client $n$ estimates the local Fisher Information for each parameter $\theta_{n,j}$. For a local dataset $D_n$, the Fisher Information for parameter $j$ is computed as:
\begin{equation}
\label{equ:fisher}
    I(\theta_{n,j}) = 
    \Bigg( \frac{\partial}{\partial \theta_{n,j}} 
    \log \mathcal{L}(\theta_n, D_n) \Bigg)^2.
\end{equation}

To ensure comparability across layers with different scales, each layer’s Fisher scores are normalized using a min-max operation. For each layer $l$, the normalized Fisher score for parameter $j$ is:
\begin{equation}
    \tilde{I}(\theta_{n,j}) = 
    \frac{I(\theta_{n,j}) - I_{\min}^l}{I_{\max}^l - I_{\min}^l},
    \quad \forall j \in l.
\end{equation}
Here, $I_{\min}^l$ and $I_{\max}^l$ denote the minimum and maximum Fisher values in layer $l$ respectively.

Finally, each client derives its local sensitive mask by thresholding the normalized scores:
\begin{equation}
\label{equ:fisher_score}
    M_n^t = \big\{ j \;|\; \tilde{I}(\theta_{n,j}) > \tau_n \big\},
\end{equation}
where $\tau_n$ is a client-specific threshold that can be tuned based on privacy preference and system capacity.

\subsection{Collaborative Secure Mask Negotiation}
\label{sec:collaborative}

To ensure accurate decryption in homomorphic encryption and support batch packing encryption, all selected clients $n \in K^t$ first compute their local sensitive masks $M_n^t$ and exchange them to establish a consistent encryption domain.

The \textbf{encrypted zone} is formally defined as the intersection of all local masks:
\begin{equation}
    M_{\text{enc}}^t = \bigcap_{n \in K^t} M_n^t.
\end{equation}
Parameters within $M_{\text{enc}}^t$ are universally recognized as highly sensitive and are consequently protected through homomorphic encryption. However, when dealing with either a large number of clients or significant heterogeneity in client data, the common intersection among clients may exhibit limited overlap. 

To address this challenge, we propose a refined definition of the \textbf{encrypted zone} by introducing a \textbf{consensus threshold hyperparameter} $\rho \in [0, 1]$. Specifically, a parameter position is incorporated into the encrypted zone $M_{\text{enc}}^t$ if and only if it is marked as important by a proportion of clients that meets or exceeds the threshold $\rho$. The definition is as follows:
\begin{equation}
\label{equ:enc_zone}
   M_{\text{enc}}^t = \left\{ j \;\middle|\; \frac{\left|\left\{ n \in K^t \mid j \in M_n^t \right\}\right|}{\left|K^t\right|} \ge \rho \right\}.
\end{equation}
This consensus mechanism ensures that the parameters protected in $M_{\text{enc}}^t$ are those widely agreed upon as highly sensitive among the client population.

The \textbf{personalized private zone} is formed by the difference between the union and the intersection:
\begin{equation}
    M_{\text{pers}}^t =  M_n^t \setminus M_{\text{enc}}^t.
\end{equation}
This region contains parameters that are sensitive for specific clients but not shared globally. These parameters remain local and are not uploaded, preserving personalization and reducing encryption overhead.
Finally, the remaining parameters constitute the \textbf{adaptive noise zone}:
\begin{equation}
    M_{\text{noise}}^t = \Theta \setminus 
    \Big( M_{\text{enc}}^t \cup M_{\text{pers}}^t \Big),
\end{equation}
where $\Theta$ denotes the full parameter set.

\subsection{Selective Homomorphic Encryption}
\label{sec:enc}
Each client $n$ computes its local model update $\Delta\theta_n^t$. The components of the update corresponding to the encrypted zone, $\{\Delta\theta_{n,j}^t\}_{j \in M_{\text{enc}}^t}$, are packed into a plaintext vector $\mathbf{v}_n^t$. This vector is then encrypted using the CKKS public key $pk$:
\begin{equation}
    \mathbf{c}_n^t = \text{Enc}(pk, \mathbf{v}_n^t).
\end{equation}
The resulting single ciphertext $\mathbf{c}_n^t$ is sent to the server $S_\text{Agg}$. According to our setup, both clients and $S_\text{Agg}$ possess only the public key but not the private key, thus they can only perform encryption and homomorphic computation operations. 
Throughout their lifecycle, client updates remain encrypted and never undergo decryption. These encrypted updates directly participate in the aggregation process while maintaining their ciphertext form.

\subsection{Adaptive Noise Addition}
\label{sec:noise}
For parameters in the adaptive noise zone $M_{\text{noise}}^t$, clients apply differential privacy using the Gaussian mechanism, analyzed under Rényi Differential Privacy (RDP). In this standard approach, each client $n$ independently and adaptively selects two parameters for round $t$: a clipping norm $C_n^t$ and a noise standard deviation $\sigma_n^t$.

First, the client clips the sub-vector of its update corresponding to the noise zone by its chosen $L_2$ clipping norm $C_n^t$:
\begin{equation}
    \bar{\Delta\theta}_{n, M_{\text{noise}}^t} = \Delta\theta_{n, M_{\text{noise}}^t} \cdot \min\left(1, \frac{C_n^t}{\|\Delta\theta_{n, M_{\text{noise}}^t}\|_2}\right).
\end{equation}
This ensures the sensitivity of the update is bounded by $C_n^t$. Then, the client adds Gaussian noise with its chosen standard deviation $\sigma_n^t$:
\begin{equation}
    \tilde{\Delta\theta}_{n,j}^t = \bar{\Delta\theta}_{n,j}^t + z_{n,j}, \quad \forall j \in M_{\text{noise}}^t,
\end{equation}
where $z_{n,j} \sim \mathcal{N}(0, (C_n^t\sigma_n^t)^2{|K^t|})$. 

For any RDP order $\alpha > 1$, this mechanism guarantees $(\alpha, \varepsilon_n(\alpha))$-RDP. According to Lemma \ref{lemma:rdp_gaussian}, the privacy budget is given by:
\begin{equation}
\varepsilon_n(\alpha) = \frac{\alpha (C_n^t)^2}{2(C_n^t\sigma_n^t)^2} = \frac{\alpha}{2(\sigma_n^t)^2}.
\end{equation}
While our framework allows each client $n$ to choose its own noise parameter $\sigma_n^t$, we focus on the uniform case $\sigma_n^t\equiv\sigma$ for all participating clients.  This simplification facilitates a global privacy analysis and guarantees a consistent, minimal level of protection.
After $T$ iterations, the mechanism satisfies $(\alpha, \frac{T\alpha}{2\sigma^2})$-RDP according to Lemma \ref{lemma:rdp_composition}. By applying Lemma \ref{lemma:rdp_conversion}, this implies $(\epsilon, \delta)$-DP after $T$ rounds, where $\epsilon = \frac{T\alpha}{2\sigma^2} + \log\left(\frac{\alpha-1}{\alpha}\right) - \frac{\log(\delta) + \log\alpha}{\alpha-1}$.

\subsection{Server Aggregation}
\label{sec:agg}
We introduce a two-server architecture to enhance security by separating key management from aggregation tasks. The system comprises a trusted \textbf{Key Distribution and Decryption Server ($S_\text{Key}$)} and a semi-trusted \textbf{Aggregation Server ($S_\text{Agg}$)}.

At the beginning of the process, $S_\text{Key}$ generates a public-secret key pair $(pk, sk)$ for the CKKS scheme. It retains the secret key $sk$ and distributes the public key $pk$ to all clients and to the Aggregation Server $S_\text{Agg}$.

The aggregation process proceeds in the following steps:

\textbf{Step 1: Aggregation at Server $S_\text{Agg}$.}
The Aggregation Server $S_\text{Agg}$ is responsible for collecting updates from all clients $n \in K^t$. Since $S_\text{Agg}$ does not possess the secret key, it performs aggregation as follows:
\begin{itemize}
    \item For the \textbf{encrypted zone} ($M_{\text{enc}}^t$), $S_\text{Agg}$ leverages the additive homomorphism of the CKKS scheme to sum the received ciphertexts without decrypting them:
    \begin{equation}
        \mathbf{c}_{\text{agg}}^t = \sum_{n \in K^t} \mathbf{c}_n^t.
    \end{equation}
    \item For the \textbf{adaptive noise zone} ($M_{\text{noise}}^t$), $S_\text{Agg}$ aggregates the noisy plaintext updates through simple summation:
    \begin{equation}
        \Delta\theta_{\text{sum},j}^t = \sum_{n \in K^t} \tilde{\Delta\theta}_{n,j}^t, \quad \forall j \in M_{\text{noise}}^t.
    \end{equation}
\end{itemize}
After completing these operations, $S_\text{Agg}$ holds a partially aggregated result, consisting of the aggregated ciphertext $\mathbf{c}_{\text{agg}}^t$ and the plaintext summed updates for the noise zone. $S_\text{Agg}$ then transmits this entire result to the Key Server $S_\text{Key}$.

\textbf{Step 2: Decryption and Finalization at Server $S_\text{Key}$.}
The trusted Key Server $S_\text{Key}$ receives the aggregated results from $S_\text{Agg}$. Using its private key $sk$, it decrypts the aggregated ciphertext:
\begin{equation}
    \mathbf{v}_{\text{sum}}^t = \text{Dec}(sk, \mathbf{c}_{\text{agg}}^t).
\end{equation}
The resulting plaintext vector $\mathbf{v}_{\text{sum}}^t$ contains the summed updates for the encrypted zone, $\{\Delta\theta_{\text{sum},j}^t\}_{j \in M_{\text{enc}}^t}$. $S_\text{Key}$ now combines this with the plaintext sums from the noise zone to form the complete aggregated update vector, $\Delta\theta_{\text{sum}}^t$.

Finally, $S_\text{Key}$ computes the new global model:
\begin{equation}
    \theta_{\text{global}}^t = \theta_{\text{global}}^{t-1} + \eta_g \frac{1}{|K^t|} \Delta\theta_{\text{sum}}^t,
\end{equation}
where $\eta_g$ is the server learning rate.
Notably, when setting $\eta_g = |K^t|$, the aggregation algorithm reduces to the standard FedAvg approach.

\textbf{Step 3: Global Model Distribution.}
$S_\text{Key}$ broadcasts the updated global model $\theta_{\text{global}}^t$ back to all clients. The clients then use this new model for their local merging step in the next round. 

Specifically, the client retains the parameters from its personalized zone ($M_{\text{pers}}^t$) that were updated during the current round of local training. For all other parameters (i.e., those in the encrypted zone $M_{\text{enc}}^t$ and the noise zone $M_{\text{noise}}^t$), it adopts the new parameters from the received global model.

This selective merge operation can be formulated as:
\begin{equation}
\theta_n^t \leftarrow M_{\text{pers}}^t \odot \theta_{n, \text{local}}^t + (1 - M_{\text{pers}}^t) \odot \theta_{\text{global}}^t
\end{equation}
where $\theta_n^t$ is the final model for client $n$ at the end of round $t$, which will be used for the next round of training, $\theta_{n, \text{local}}^t$ is the model of client $n$ after completing local training in round $t$, $\theta_{\text{global}}^t$ is the latest global model broadcast from the server, $M_{\text{pers}}^t$ is the personalized parameter mask for client $n$, $\odot$ represents the Hadamard product (element-wise multiplication).
This merged model, which contains both global consensus and personalized knowledge, is then used by the client to begin the next training round.

\section{Evaluation}
\subsection{Experimental Setup}

We conduct comprehensive experiments on five datasets under various conditions to evaluate the performance of the proposed method in comparison with existing approaches. Specifically, we assess the accuracy on these datasets, comparing our SelectiveShield method (abbreviated as "Ours" in subsequent tables for conciseness) against DP-FedAvg, MaskCrypt, FedML-HE, and DPSGD.

\textbf{Datasets and Models.} We use five benchmark datasets in federated learning and privacy research: MNIST ~\cite{lecun1998gradient}, Fashion-MNIST (FMNIST) ~\cite{xiao2017fashion}, CIFAR-10 ~\cite{krizhevsky2009learning}, CIFAR-100 ~\cite{krizhevsky2009learning}, and SVHN ~\cite{netzer2011reading}. These datasets vary in complexity, ranging from simple grayscale digits (MNIST) to diverse real-world images (SVHN). 

For MNIST and FMNIST datasets, we employ a 3-layer MLP architecture. For other datasets, we utilize a CNN model consisting of 2 convolutional layers followed by 2 fully-connected layers.

\textbf{Comparison Methods.} We evaluate the performance of our proposed method against several state-of-the-art approaches in differentially private federated learning, including DP-FedAvg~\cite{geyer2017differentially}, DPSGD~\cite{abadi2016deep}, FedML-HE~\cite{jin2023fedmlhe}, and MaskCrypt~\cite{hu2024maskcrypt}. Notably, DP-FedAvg implements user-level differential privacy, while DPSGD provides record-level privacy guarantees. Both FedML-HE and MaskCrypt employ selective encryption techniques for privacy preservation.

\textbf{Attacks.} We evaluate SelectiveShield against {Gradient Inversion Attacks (GIA)}. We use iDLG~\cite{zhao2020idlg} to reconstruct input samples from gradients.

\textbf{Implementation Details.} For all datasets, we set the learning rate to 0.01. 
We employ 20 clients with 10 global epochs, 5 local epochs per client, and a batch size of 32.
We use the Rényi Differential Privacy (RDP) algorithm provided by Opacus as our privacy accountant. 
After performing a grid search over $\alpha$, $\tau$, and $\rho$, we adopted the parameter combination that achieved the best validation performance; the complete search results are provided in the appendix for reference.

We employ a Dirichlet-distributed partition to synthesize \emph{non-IID} data, controlled by the concentration parameter $\alpha$: as $\alpha\to0$, clients possess highly heterogeneous class distributions; increasing $\alpha$ approaches IID. 
All experiments were implemented in Python with PyTorch on an NVIDIA 4090 GPUs and an Intel Xeon Platinum 8369B CPU.

\subsection{Performance Comparison}

We evaluate the model accuracy of various methods on five datasets under statistically heterogeneous scenarios with Dirichlet distribution ($\alpha=0.5$) and a privacy budget of $\epsilon=1.0$. Our comparative analysis between SelectiveShield and baseline methods, as presented in Table~\ref{tab:performance_1} and Table~\ref{tab:performance_2}, reveals that our approach consistently delivers competitive or superior performance across all evaluated datasets.
\begin{table}[htbp]
  \centering
    \begin{tabular}{ccc}
    \toprule
    \textbf{Algorithm} & \textbf{CIFAR10} & \textbf{CIFAR100} \\
    \midrule
    \midrule
    DP-FedAvg & 0.538  & 0.245  \\
    MaskCrypt & 0.566  & 0.329  \\
    FedML-HE & 0.571  & 0.270  \\
    DPSGD & 0.492  & 0.308  \\
    Ours($\tau=0.2$) & \textbf{0.587}  & 0.397  \\
    Ours($\tau=0.3$) & 0.542  & \textbf{0.416}  \\
    Ours($\tau=0.5$) & 0.573  & 0.415  \\
    \bottomrule
    \end{tabular}%
  \caption{Performance Comparison} \label{tab:performance_1}
\end{table}%
\begin{table}[htbp]
  \centering
    \begin{tabular}{cccc}
    \toprule
    \textbf{Algorithm} & \textbf{MNIST} & \textbf{FMNIST} & \textbf{SVHN} \\
    \midrule
    \midrule
    DP-FedAvg & 0.982  & 0.884  & 0.873  \\
    MaskCrypt & 0.981  & 0.887  & 0.867  \\
    FedML-HE & 0.982  & 0.889  & 0.877  \\
    DPSGD & \textbf{0.986}  & 0.872  & 0.846  \\
    Ours($\tau=0.02$) & 0.981  & 0.892  & 0.877  \\
    Ours($\tau=0.05$) & 0.982 & \textbf{0.894} & \textbf{0.880} \\
    \bottomrule
    \end{tabular}%
  \caption{Performance Comparison}\label{tab:performance_2}
\end{table}%

\subsection{Impact of Data Heterogeneity}
To evaluate the robustness of SelectiveShield under non-IID conditions, we conducted experiments on CIFAR-10 with varying degrees of data heterogeneity, controlled by the Dirichlet distribution parameter $\alpha$. As shown in Table~\ref{tab:acc_vs_noiid}, our method demonstrates strong performance across different levels of non-IID data. 
When data heterogeneity is high ($\alpha=0.2$ and $\alpha=0.5$), SelectiveShield ($\tau=0.5$ and $\tau=0.2$ respectively) achieves the highest accuracy. In the IID case, our method still achieves the best performance.
\begin{table}[htbp]
  \centering
    \begin{tabular}{ccccc}
    \toprule
    \multirow{2}[4]{*}{\textbf{Algorithm}} & \multicolumn{3}{c}{\textbf{Dirichlet $\alpha$}} & \multirow{2}[4]{*}{\textbf{iid}} \\
\cmidrule{2-4}          & \textbf{0.2} & \textbf{0.5} & \textbf{1} &  \\
    \midrule
    \midrule
    DP-FedAvg & 0.649  & 0.538  & 0.744  & 0.745  \\
    MaskCrypt & 0.642  & 0.566  & 0.748  & 0.759  \\
    FedML-HE & 0.644  & 0.571  & 0.747  & 0.749  \\
    PPSGD & 0.594  & 0.236  & 0.662  & 0.758  \\
    Ours($\tau=0.2$) & 0.648  & \textbf{0.587}  & \textbf{0.752}  & 0.757  \\
    Ours($\tau=0.3$) & 0.653  & 0.542  & 0.751  & 0.760  \\
    Ours($\tau=0.5$) & \textbf{0.655}  & 0.573  & 0.740  & \textbf{0.764}  \\
    \bottomrule
    \end{tabular}%
  \caption{Accuracy vs. non-IID Degree on CIFAR-10}\label{tab:acc_vs_noiid}
\end{table}%

\subsection{Hyperparameter Analysis}

Our proposed method adaptively balances the ratio of parameters that are encrypted, personalized, and perturbed with noise, ensuring that all parameters receive a degree of protection. For instance, if the scope of both the encrypted and personalized zones approaches zero, all parameters will have noise added, causing our framework to degenerate into a standard local differential privacy (DP) approach in federated learning. Therefore, a detailed analysis of strengthening privacy by increasing the privacy budget (i.e., lowering $\epsilon$) is not the primary focus of this section. Since the encrypted and personalized parameters do not leak any information to an attacker, the system remains secure as long as the privacy budget for the remaining parameters is sufficiently high.

Consequently, in this section, we conduct an ablation study to evaluate the resilience of our framework when the noise-adding module is removed. We test the effectiveness of the encryption and personalization components alone against the iLRG (instance-level label restoration from gradients) inversion attack. The attack's performance is measured using two metrics: (1) Label Existence Accuracy (LeAcc), which is the accuracy score for predicting the existence of a label, and (2) Label Number Accuracy (LnAcc), which is the accuracy score for predicting the number of instances per class.

\textbf{Effect of Sensitivity Threshold ($\tau$):}
The threshold $\tau$ controls the size of each client's sensitive parameter mask. As $\tau$ increases from 0.01 to 1.0, the encrypted zone ($M_{\text{enc}}$) shrinks from 28.64\% to 0\%. This directly impacts security: the attacker's LeAcc climbs from an ineffective 0.4 to a perfect 1.0. This demonstrates that $\tau$ is a critical lever for managing the trade-off between computational efficiency and security.

\textbf{Effect of Consensus Threshold ($\rho$):}
The consensus threshold $\rho$ sets the requirement for a parameter to be included in the globally encrypted zone. Increasing $\rho$ from 0.1 to 0.7 tightens this consensus, reducing the encrypted zone from 5.52\% to 1.06\%. While a smaller encrypted zone leads to a high LeAcc, the attack's Label Number Accuracy (LnAcc) remains low. This shows that even a small, strategically chosen encrypted zone is effective at preventing the attacker from learning the exact number of data instances per class. Therefore, $\rho$ fine-tunes the balance between a unified global defense and client-specific personalization.

The detailed results presented in the appendix demonstrate comprehensive experiments conducted across various combinations of parameters $\alpha$, $\tau$, and $\rho$. These experiments systematically evaluate their impacts on multiple performance metrics, including: classification accuracy, the effective ranges of encryption masks ($M_\textbf{enc}$), personalization masks ($M_\textbf{pers}$), and noise masks ($M_\textbf{noise}$), as well as the computational time costs associated with encryption, decryption, aggregation, and training processes.

\begin{table}[]
\centering
\begin{tabular}{@{}cccccc@{}}
\toprule
 & $\tau$ & $M_{\text{enc}}$ & $M_{\text{pers}}$  & LeAcc  & LnAcc \\ \midrule
& 0.01   & 28.64\%   & 13.19\%  & 0.4  & 0.0   \\
& 0.02   & 19.68\%   & 11.29\%  & 0.2  & 0.0   \\
& 0.07   & 4.89\%    & 0.98\%   & 0.4  & 0.0   \\
& 0.1    & 2.28\%    & 1.00\%   & 0.9  & 0.1   \\
& 0.2    & 0.45\%    & 0.27\%   & 0.9  & 0.2   \\
& 1.0    & 0.00\%    & 0.00\%   & 1.0  & 0.6   \\ \bottomrule
\end{tabular}%
\caption{Hyperparameter Study on $\tau$ Refer Eq. (\ref{equ:fisher_score})}
\label{tab:tau}
\end{table}
\begin{table}[]
\centering
\begin{tabular}{@{}ccccc@{}}
\toprule
$\rho$ & $M_{\text{enc}}$ & $M_{\text{pers}}$  & LeAcc  & LnAcc\\ \midrule
0.1    & 5.52\%   & 3.94\%   & 0.5  & 0.0 \\
0.2    & 4.13\%   & 1.65\%   & 1.0  & 0.2 \\
0.3    & 4.24\%   & 2.41\%   & 0.9  & 0.1 \\
0.5    & 2.28\%   & 1.00\%   & 0.9  & 0.1 \\
0.7    & 1.06\%   & 0.54\%   & 1.0  & 0.3 \\
0.8    & 1.35\%   & 0.69\%   & 0.9  & 0.2 \\
\bottomrule
\end{tabular}
\caption{Hyperparameter Study on $\rho$ Refer Eq. (\ref{equ:enc_zone})}
\label{tab:rho}
\end{table}

\section{Conclusion}
In this paper, we introduced SelectiveShield, a lightweight and adaptive hybrid defense framework designed to address the challenges of privacy in heterogeneous federated learning. Our framework leverages Fisher information and a collaborative negotiation protocol to partition parameters into unique encrypted, personalized, and noise-protected zones. This partitioning method, inspired by personalized federated learning, has been demonstrated through extensive experiments to maintain high model utility under non-independent and identically distributed (non-IID) scenarios. Future work will focus on optimizing the communication protocol and evaluating the framework against more advanced threat models.

\bibliography{aaai2026}

\clearpage
\appendix
\onecolumn 

\section{Notations}
\label{sec:notations}

This section defines the key notations and hyperparameters used throughout the paper and this supplementary material. A summary is provided in Table \ref{tab:notations}.

\begin{table}[h!]
\centering
\begin{tabular}{@{}ll@{}}
\textbf{Symbol} & \textbf{Description} \\
\midrule
$\theta$ & Model parameters. \\
$T$ & Total communication rounds. \\
$K^t$ & Set of participating clients in round $t$. \\
$D_n$ & Local dataset of client $n$. \\
$\Delta\theta_n^t$ & Local model update from client $n$ at round $t$. \\
$\alpha$ & Controls data heterogeneity (non-IID) via Dirichlet distribution. \\
$\tau$ & Fisher Information threshold for selecting sensitive parameters. \\
$\rho$ & Consensus threshold for defining the encrypted zone. \\
$\eta_g$ & Server-side global learning rate. \\
$I(\theta_{n,j})$ & Fisher Information value for a specific parameter. \\
$M_n^t$ & Local mask of sensitive parameters for client $n$. \\
$M_{\text{enc}}$ & Mask for the encrypted zone (protected by Homomorphic Encryption). \\
$M_{\text{pers}}$ & Mask for the personalized zone (parameters retained locally). \\
$M_{\text{noise}}$ & Mask for the adaptive noise zone (protected by Differential Privacy). \\
$pk, sk$ & Public and secret key pair for Homomorphic Encryption. \\
$\mathbf{c}_n^t$ & Ciphertext of a client's update. \\
$C_n^t$ & $L_2$ norm clipping bound for Differential Privacy. \\
$\sigma_n^t$ & Standard deviation of Gaussian noise for Differential Privacy. \\
$S_\text{Key}$ & Key Distribution and Decryption Server (KDS). \\
$S_\text{Agg}$ & Aggregation Server (AS). \\
\bottomrule
\end{tabular}
\caption{List of Notations and Hyperparameters.}
\label{tab:notations}
\end{table}

\section{Experimental Setup Details}

\subsection{Model Architectures}
\label{subsec:model_architectures}

For reproducibility, we provide the detailed architectures of the models used in our experiments. For the MNIST and Fashion-MNIST datasets, a 3-layer Multilayer Perceptron (MLP) was used. For the CIFAR-10, CIFAR-100, SVHN and stl10 datasets, a Convolutional Neural Network (CNN) was employed. The specifics of these architectures, based on the implementation code, are detailed in Table \ref{tab:model_architectures}. Parameters that vary by dataset, such as input or output dimensions, are represented by variable names.

\begin{table}[h!]
\centering
\begin{tabular}{@{}ll@{}}
$\to$prule
\textbf{Dataset} & \textbf{Model Architecture} \\ \midrule
\multirow{4}{*}{MNIST / FMNIST} & \textbf{MLP} \\
 & Linear (input\_dim, 256) $\to$ ReLU \\
 & Linear (256, 128) $\to$ ReLU \\
 & Linear (128, num\_classes) \\ \midrule
\multirow{6}{*}{\shortstack{CIFAR-10 / CIFAR-100 \\ SVHN / stl10}} & \textbf{SimpleCNN} \\
 & Conv2d (in\_channels, 32, kernel\_size=3, padding=1) $\to$ BatchNorm2d $\to$ ReLU $\to$ MaxPool2d(2) \\
 & Conv2d (32, 64, kernel\_size=3, padding=1) $\to$ BatchNorm2d $\to$ ReLU $\to$ MaxPool2d(2) \\
 & Flatten \\
 & Linear (n\_features, 256) $\to$ ReLU \\
 & Linear (256, num\_classes) \\ \bottomrule
\end{tabular}
\caption{Model architectures used in the experiments.}
\label{tab:model_architectures}
\end{table}

\subsection{Data Partitioning}

\begin{figure*}[h!]
\centering
\includegraphics[width=0.55\linewidth]{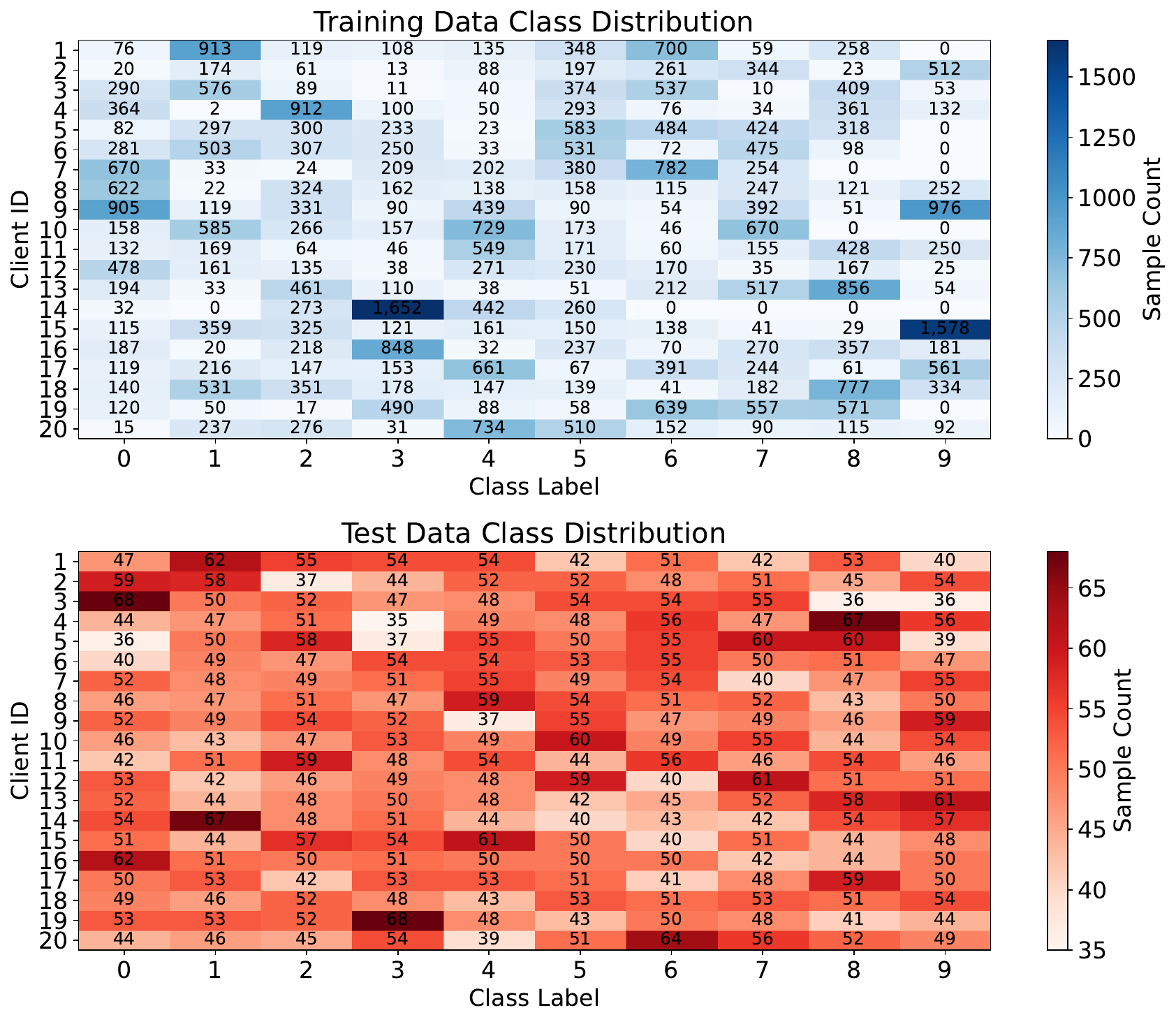}
\caption{Client-wise class distributions with $\alpha=1$ (non-IID).}
\label{fig:data_partition_alpha1}
\end{figure*}

\begin{figure*}[h!]
\centering
\includegraphics[width=0.55\linewidth]{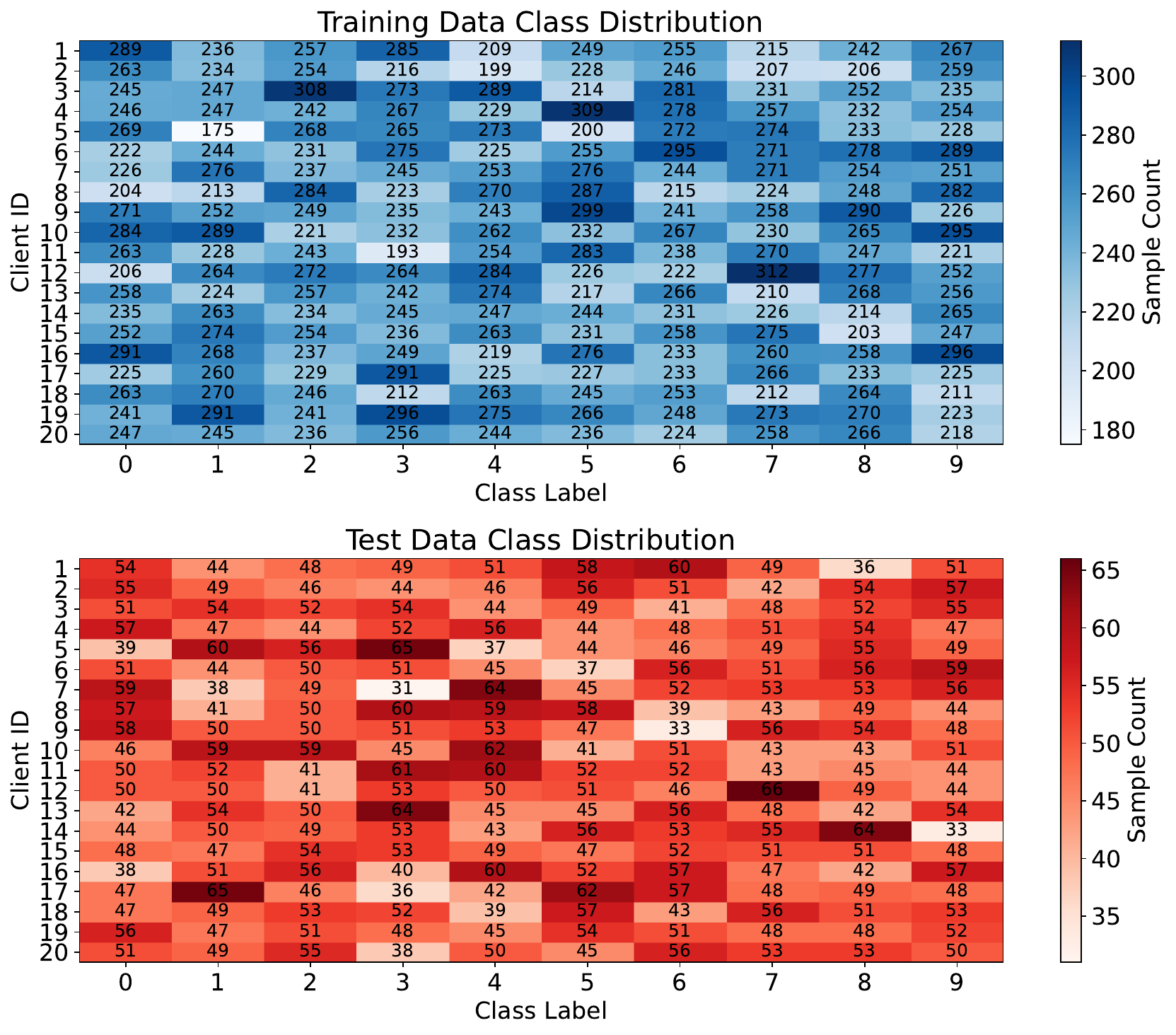}
\caption{Client-wise class distributions with $\alpha=100$ (more IID).}
\label{fig:data_partition_alpha100}
\end{figure*}

To simulate non-IID data distributions, we partitioned the datasets among 20 clients using a Dirichlet distribution, controlled by the concentration parameter $\alpha$. A smaller $\alpha$ results in higher heterogeneity, where clients possess distinct class distributions, as illustrated in Figure \ref{fig:data_partition_alpha1}. In contrast, a larger $\alpha$ yields more IID-like distributions across clients, as shown in Figure \ref{fig:data_partition_alpha100}.

\section{Fisher Information Visualization}

The core of SelectiveShield is identifying sensitive parameters using Fisher Information (FI). This process involves three key steps: calculating raw FI values, normalizing them for comparability, and applying a threshold to generate a sensitivity mask. The following figures and formulas illustrate this process.

\subsection{Step 1: Raw Fisher Information Calculation}
\begin{figure*}[h!]
    \centering
    \includegraphics[width=1\linewidth]{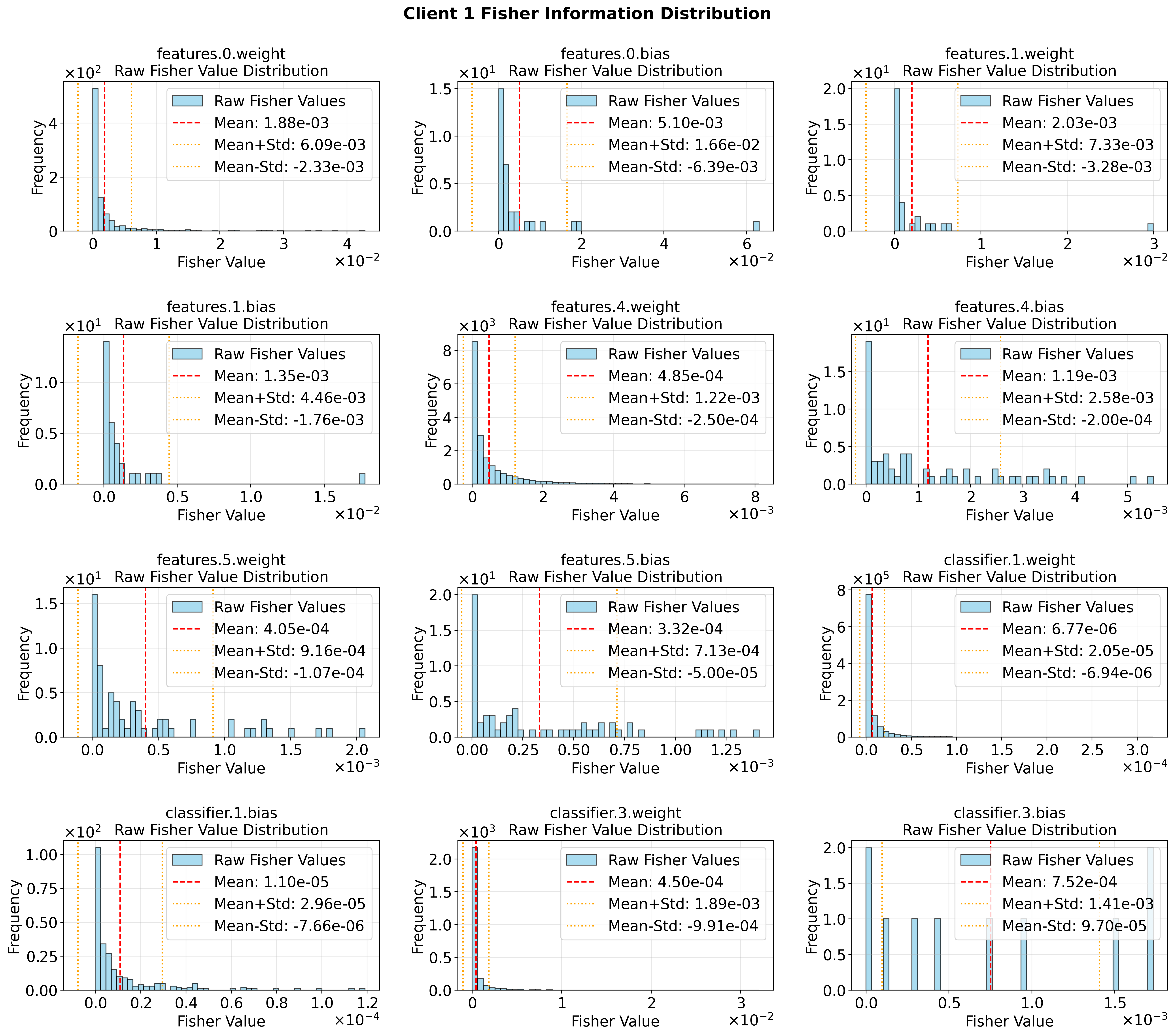}
    \caption{Raw Fisher Information Distribution.}
    \label{fig:raw_fisher_dist}
\end{figure*}

First, each client computes the local Fisher Information for every parameter $\theta_{n,j}$. This value, which approximates the parameter's importance to the model's loss, is calculated as the squared gradient of the log-likelihood, as shown in Equation \ref{equ:fisher_calc}.
\begin{equation}
    \label{equ:fisher_calc}
    I(\theta_{n,j}) = 
    \Bigg( \frac{\partial}{\partial \theta_{n,j}} 
    \log \mathcal{L}(\theta_n, D_n) \Bigg)^2
\end{equation}
Figure \ref{fig:raw_fisher_dist} visualizes the distribution of these raw FI values for a single client. The histograms show that the distributions are typically heavy-tailed, with most parameters having very small FI values and a few having very high values. This supports our strategy of selectively protecting a small subset of important parameters.

\subsection{Step 2: Normalization}
\begin{figure*}[h!]
    \centering
    \includegraphics[width=0.85\linewidth]{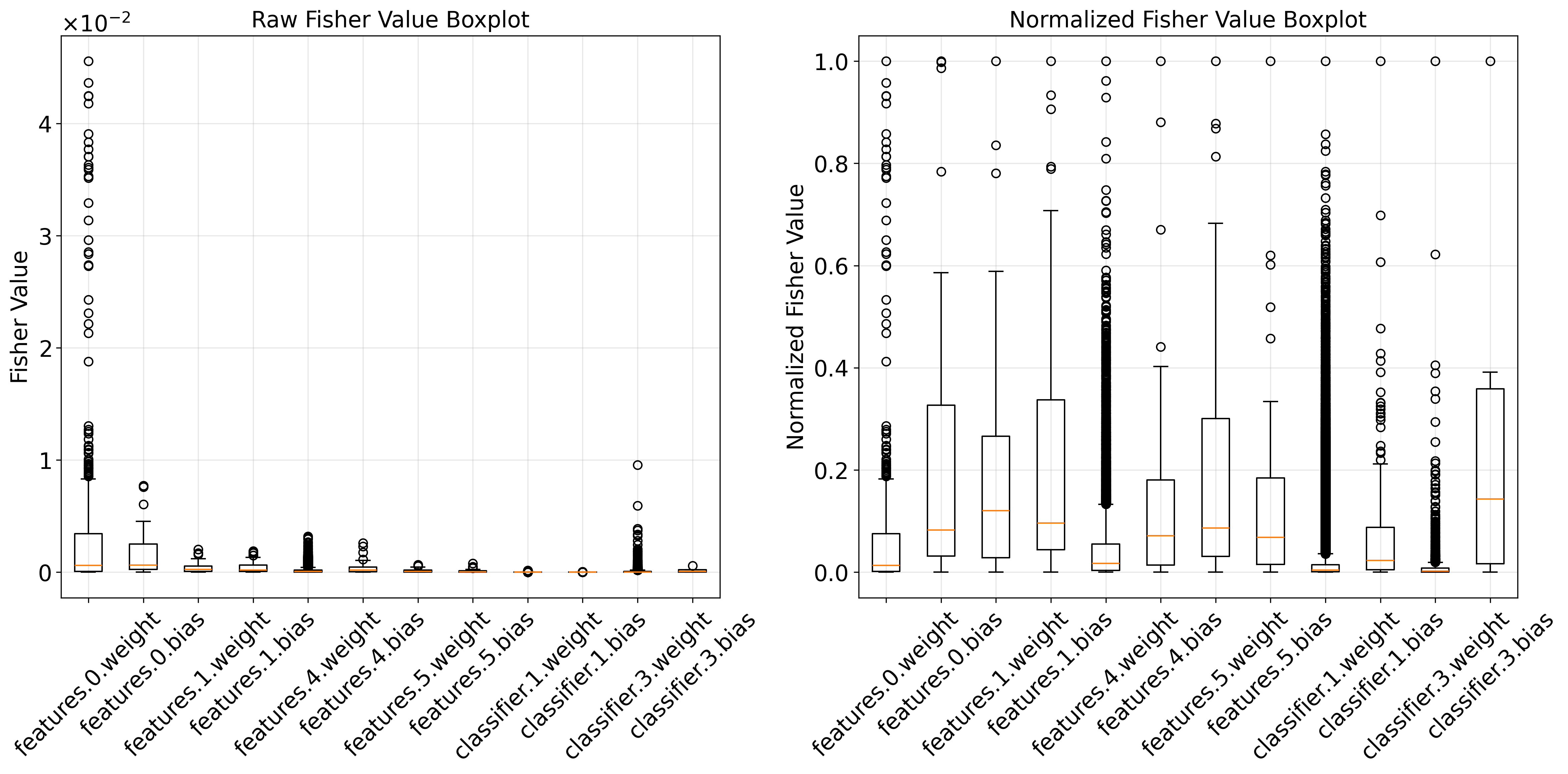}
    \caption{Fisher Value Boxplots (Raw and Normalized).}
   \label{fig:fisher_value_boxplot}
\end{figure*}

As shown in the left plot of Figure \ref{fig:fisher_value_boxplot}, the raw FI values can have vastly different scales across different model layers. This makes it difficult to apply a single, meaningful threshold to the entire model. To address this, we apply a min-max normalization to the FI scores within each layer $l$, as defined in Equation \ref{equ:fisher_norm}.
\begin{equation}
    \label{equ:fisher_norm}
    \tilde{I}(\theta_{n,j}) = 
    \frac{I(\theta_{n,j}) - I_{\min}^l}{I_{\max}^l - I_{\min}^l},
    \quad \forall j \in l
\end{equation}
Here, $I_{\min}^l$ and $I_{\max}^l$ denote the minimum and maximum Fisher values in layer $l$, respectively. The right plot in Figure \ref{fig:fisher_value_boxplot} shows the result of this normalization, where all values are brought to a comparable scale of [0, 1].

\subsection{Step 3: Thresholding and Mask Generation}
\begin{figure*}[h!]
    \centering
    \includegraphics[width=1\linewidth]{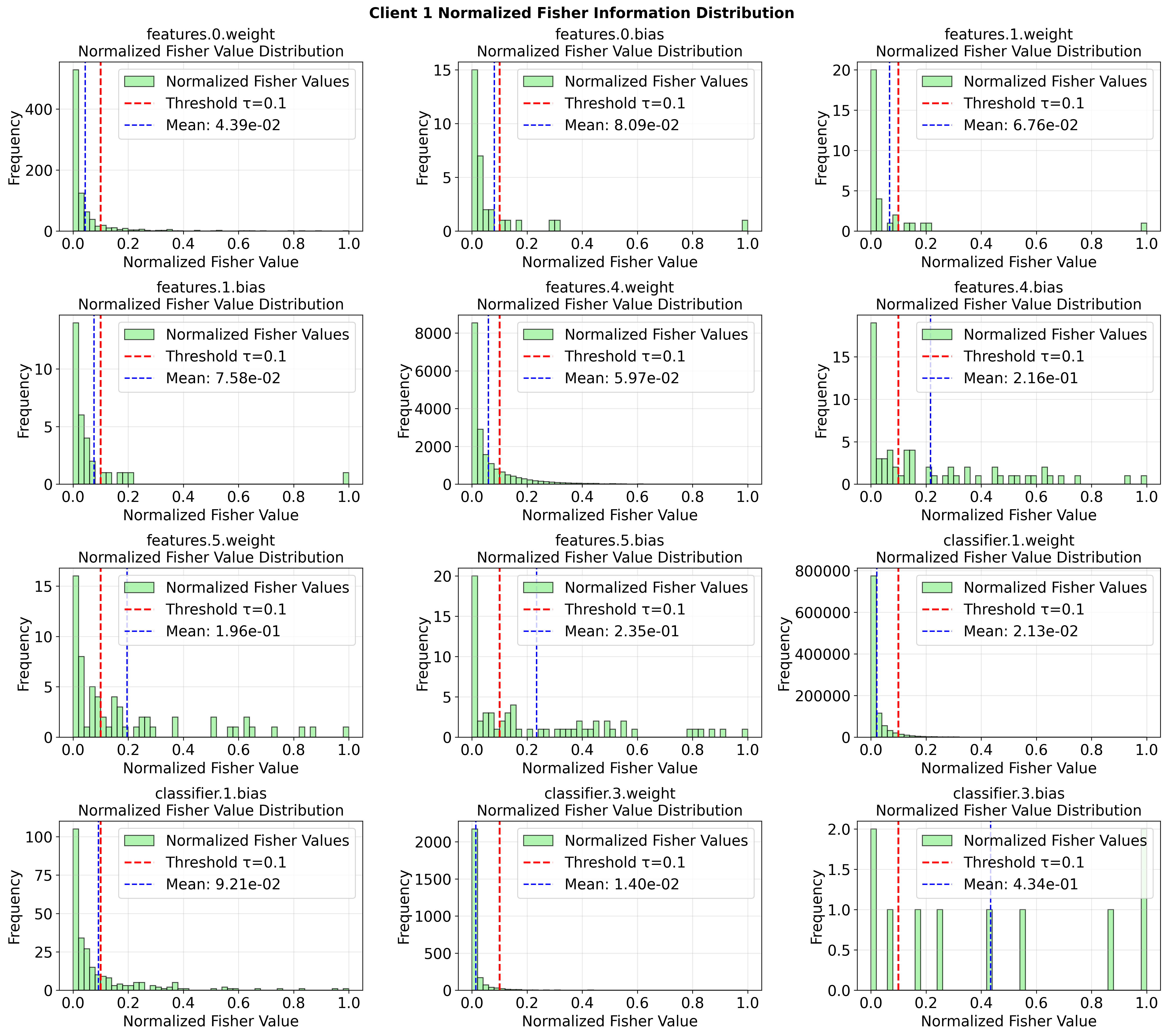}
    \caption{Normalized Fisher Information Distribution.}
    \label{fig:norm_fisher_dist}
\end{figure*}

With the normalized FI scores, each client can now derive its local sensitive mask, $M_n^t$, by selecting all parameters whose normalized FI score exceeds a client-specific threshold $\tau_n$.
\begin{equation}
    \label{equ:fisher_mask}
    M_n^t = \big\{ j \;|\; \tilde{I}(\theta_{n,j}) > \tau_n \big\}
\end{equation}
Figure \ref{fig:norm_fisher_dist} provides a clear visualization of this step. It shows the distribution of the normalized FI scores, with a vertical dashed line representing a sample threshold ($\tau=0.1$). This demonstrates how the threshold effectively separates the more sensitive parameters (to the right of the line) from the less sensitive ones, finalizing the local selection process.

\clearpage

\section{Detailed Experimental Results}
This section provides the complete results from our extensive hyperparameter search, as mentioned in the main paper. The following tables detail the performance (Accuracy), mask ratios ($M_{\text{enc}}$, $M_{\text{pers}}$, $M_{\text{noise}}$), and time costs for different combinations of the Dirichlet parameter $\alpha$, sensitivity threshold $\tau$, and consensus threshold $\rho$ across all tested datasets.
\begin{table}[htbp]
  \centering
    \begin{tabular}{ccccccccccc}
    \toprule
    \multicolumn{3}{c}{\textbf{Hyperparameters}} & \multicolumn{3}{c}{\textbf{Mask ratio}} & \multicolumn{4}{c}{\textbf{Time Cost (s)}} & \textbf{Performance} \\
    \midrule
    \midrule
    $\alpha$  & $\tau$  & $\rho$  &  $M_{\text{enc}}$ &  $M_{\text{pers}}$ &  $M_{\text{noise}}$ & Encryption & Decryption & Aggregation & Training & Acc \\
    \midrule
    \multirow{9}[6]{*}{0.2} & \multirow{3}[2]{*}{0.05} & 0.3   & 3.56\% & 3.27\% & 93.17\% & 3.25  & 0.04  & 0.95  & 6.96  & 0.980  \\
          &       & 0.5   & 1.47\% & 2.79\% & 95.74\% & 3.22  & 0.04  & 0.94  & 6.97  & 0.979  \\
          &       & 0.7   & 0.46\% & 2.94\% & 96.60\% & 2.71  & 0.03  & 0.80  & 7.01  & 0.982  \\
\cmidrule{2-11}          & \multirow{3}[2]{*}{0.1} & 0.3   & 1.33\% & 1.54\% & 97.13\% & 3.23  & 0.04  & 0.94  & 7.29  & 0.980  \\
          &       & 0.5   & 0.42\% & 1.16\% & 98.42\% & 3.28  & 0.04  & 0.94  & 7.14  & 0.980  \\
          &       & 0.7   & 0.11\% & 1.16\% & 98.73\% & 2.70  & 0.03  & 0.79  & 7.07  & 0.981  \\
\cmidrule{2-11}          & \multirow{3}[2]{*}{0.2} & 0.3   & 0.47\% & 0.64\% & 98.89\% & 3.27  & 0.04  & 0.94  & 7.04  & 0.981  \\
          &       & 0.5   & 0.12\% & 0.46\% & 99.42\% & 3.01  & 0.04  & 0.88  & 7.09  & 0.981  \\
          &       & 0.7   & 0.01\% & 0.45\% & 99.53\% & 1.88  & 0.02  & 0.56  & 6.84  & 0.980  \\
    \midrule
    \multirow{9}[6]{*}{0.5} & \multirow{3}[2]{*}{0.05} & 0.3   & 5.08\% & 4.96\% & 89.96\% & 3.25  & 0.04  & 0.94  & 5.94  & 0.981  \\
          &       & 0.5   & 1.93\% & 4.33\% & 93.73\% & 3.25  & 0.04  & 0.97  & 5.86  & 0.981  \\
          &       & 0.7   & 0.19\% & 4.27\% & 95.54\% & 3.15  & 0.04  & 0.92  & 5.89  & 0.981  \\
\cmidrule{2-11}          & \multirow{3}[2]{*}{0.1} & 0.3   & 2.25\% & 2.57\% & 95.19\% & 3.23  & 0.04  & 0.93  & 5.83  & 0.980  \\
          &       & 0.5   & 0.27\% & 1.77\% & 97.96\% & 3.24  & 0.04  & 0.95  & 6.15  & 0.982  \\
          &       & 0.7   & 0.02\% & 1.86\% & 98.11\% & 2.60  & 0.03  & 0.86  & 5.91  & 0.983  \\
\cmidrule{2-11}          & \multirow{3}[2]{*}{0.2} & 0.3   & 0.46\% & 0.76\% & 98.78\% & 3.32  & 0.04  & 1.01  & 6.08  & 0.981  \\
          &       & 0.5   & 0.04\% & 0.65\% & 99.31\% & 3.19  & 0.04  & 0.91  & 5.80  & 0.981  \\
          &       & 0.7   & 0.00\% & 0.57\% & 99.43\% & 0.75  & 0.01  & 0.27  & 5.92  & 0.982  \\
    \midrule
    \multirow{9}[6]{*}{1} & \multirow{3}[2]{*}{0.05} & 0.3   & 5.08\% & 4.51\% & 90.41\% & 3.26  & 0.04  & 0.93  & 5.72  & 0.982  \\
          &       & 0.5   & 1.98\% & 4.10\% & 93.92\% & 3.35  & 0.04  & 0.94  & 5.64  & 0.980  \\
          &       & 0.7   & 0.49\% & 4.37\% & 95.14\% & 3.27  & 0.04  & 0.96  & 5.66  & 0.981  \\
\cmidrule{2-11}          & \multirow{3}[2]{*}{0.1} & 0.3   & 1.66\% & 2.00\% & 96.34\% & 3.32  & 0.04  & 0.95  & 5.63  & 0.981  \\
          &       & 0.5   & 0.50\% & 1.83\% & 97.67\% & 3.26  & 0.04  & 0.93  & 5.80  & 0.981  \\
          &       & 0.7   & 0.03\% & 1.70\% & 98.27\% & 3.10  & 0.04  & 0.90  & 5.74  & 0.981  \\
\cmidrule{2-11}          & \multirow{3}[2]{*}{0.2} & 0.3   & 0.30\% & 0.61\% & 99.09\% & 3.35  & 0.04  & 1.03  & 5.71  & 0.981  \\
          &       & 0.5   & 0.03\% & 0.50\% & 99.48\% & 3.35  & 0.04  & 0.97  & 5.67  & 0.982  \\
          &       & 0.7   & 0.01\% & 0.62\% & 99.38\% & 1.49  & 0.02  & 0.49  & 5.65  & 0.981  \\
    \bottomrule
    \end{tabular}%
  \caption{Experimental Results on MNIST.}
  \label{tab:ref_mnist}%
\end{table}%
\begin{table}[htbp]
  \centering
    \begin{tabular}{ccccccccccc}
    \toprule
    \multicolumn{3}{c}{\textbf{Hyperparameters}} & \multicolumn{3}{c}{\textbf{Mask ratio}} & \multicolumn{4}{c}{\textbf{Time Cost (s)}} & \textbf{Performance} \\
    \midrule
    \midrule
    $\alpha$  & $\tau$  & $\rho$  &  $M_{\text{enc}}$ &  $M_{\text{pers}}$ &  $M_{\text{noise}}$ & Encryption & Decryption & Aggregation & Training & Acc \\
    \midrule
    \multirow{9}[6]{*}{0.2} & \multirow{3}[2]{*}{0.05} & 0.3   & 20.57\% & 15.70\% & 63.73\% & 4.34  & 0.05  & 1.21  & 6.66  & 0.853  \\
          &       & 0.5   & 9.28\% & 15.45\% & 75.26\% & 3.40  & 0.04  & 1.05  & 6.83  & 0.893  \\
          &       & 0.7   & 1.07\% & 17.57\% & 81.36\% & 3.24  & 0.04  & 0.96  & 6.80  & 0.877  \\
\cmidrule{2-11}          & \multirow{3}[2]{*}{0.1} & 0.3   & 10.22\% & 10.30\% & 79.48\% & 3.73  & 0.05  & 1.04  & 6.76  & 0.872  \\
          &       & 0.5   & 2.53\% & 9.25\% & 88.22\% & 3.20  & 0.04  & 0.93  & 6.67  & 0.893  \\
          &       & 0.7   & 0.18\% & 8.93\% & 90.89\% & 2.80  & 0.04  & 0.89  & 6.99  & 0.887  \\
\cmidrule{2-11}          & \multirow{3}[2]{*}{0.2} & 0.3   & 2.63\% & 4.51\% & 92.86\% & 3.44  & 0.04  & 0.96  & 6.68  & 0.893  \\
          &       & 0.5   & 0.05\% & 3.21\% & 96.74\% & 3.03  & 0.04  & 0.89  & 6.99  & 0.894  \\
          &       & 0.7   & 0.00\% & 3.86\% & 96.14\% & 0.55  & 0.01  & 0.20  & 6.74  & 0.892  \\
    \midrule
    \multirow{9}[6]{*}{0.5} & \multirow{3}[2]{*}{0.05} & 0.3   & 19.63\% & 15.34\% & 65.03\% & 5.21  & 0.06  & 1.50  & 7.03  & 0.892  \\
          &       & 0.5   & 10.07\% & 14.56\% & 75.36\% & 3.80  & 0.05  & 1.08  & 7.06  & 0.886  \\
          &       & 0.7   & 2.57\% & 16.24\% & 81.20\% & 3.24  & 0.04  & 0.94  & 7.27  & 0.889  \\
\cmidrule{2-11}          & \multirow{3}[2]{*}{0.1} & 0.3   & 10.87\% & 10.46\% & 78.67\% & 3.78  & 0.05  & 1.11  & 7.28  & 0.893  \\
          &       & 0.5   & 4.31\% & 8.44\% & 87.25\% & 3.24  & 0.04  & 0.93  & 7.26  & 0.887  \\
          &       & 0.7   & 0.01\% & 8.59\% & 91.40\% & 3.27  & 0.04  & 0.94  & 7.08  & 0.888  \\
\cmidrule{2-11}          & \multirow{3}[2]{*}{0.2} & 0.3   & 4.23\% & 5.09\% & 90.68\% & 3.24  & 0.04  & 0.95  & 7.02  & 0.889  \\
          &       & 0.5   & 0.88\% & 4.17\% & 94.95\% & 3.30  & 0.04  & 0.95  & 7.19  & 0.890  \\
          &       & 0.7   & 0.08\% & 3.99\% & 95.93\% & 2.51  & 0.03  & 0.76  & 7.16  & 0.894  \\
    \midrule
    \multirow{9}[6]{*}{1} & \multirow{3}[2]{*}{0.05} & 0.3   & 21.16\% & 14.69\% & 64.14\% & 5.19  & 0.06  & 1.48  & 6.29  & 0.891  \\
          &       & 0.5   & 11.84\% & 14.52\% & 73.64\% & 3.64  & 0.04  & 0.99  & 6.07  & 0.886  \\
          &       & 0.7   & 2.31\% & 18.58\% & 79.11\% & 3.33  & 0.04  & 0.97  & 6.04  & 0.897  \\
\cmidrule{2-11}          & \multirow{3}[2]{*}{0.1} & 0.3   & 11.95\% & 10.66\% & 77.38\% & 3.72  & 0.05  & 1.07  & 6.02  & 0.884  \\
          &       & 0.5   & 1.81\% & 8.51\% & 89.68\% & 3.20  & 0.04  & 0.92  & 6.37  & 0.887  \\
          &       & 0.7   & 0.43\% & 9.67\% & 89.90\% & 3.24  & 0.04  & 0.94  & 6.08  & 0.896  \\
\cmidrule{2-11}          & \multirow{3}[2]{*}{0.2} & 0.3   & 3.63\% & 4.48\% & 91.89\% & 3.29  & 0.04  & 0.98  & 6.00  & 0.888  \\
          &       & 0.5   & 0.62\% & 4.23\% & 95.15\% & 3.24  & 0.04  & 0.93  & 6.06  & 0.887  \\
          &       & 0.7   & 0.00\% & 3.60\% & 96.40\% & 2.32  & 0.03  & 0.68  & 6.31  & 0.893  \\
    \bottomrule
    \end{tabular}%
  \caption{Experimental Results on FMNIST.}
  \label{tab:ref_fmnist}%
\end{table}%
\begin{table}[htbp]
  \centering
    \begin{tabular}{ccccccccccc}
    \toprule
    \multicolumn{3}{c}{\textbf{Hyperparameters}} & \multicolumn{3}{c}{\textbf{Mask ratio}} & \multicolumn{4}{c}{\textbf{Time Cost (s)}} & \textbf{Performance} \\
    \midrule
    \midrule
    $\alpha$  & $\tau$  & $\rho$  &  $M_{\text{enc}}$ &  $M_{\text{pers}}$ &  $M_{\text{noise}}$ & Encryption & Decryption & Aggregation & Training & Acc \\
    \midrule
    \multirow{12}[8]{*}{0.2} & \multirow{3}[2]{*}{0.2} & 0.3   & 16.14\% & 12.41\% & 71.45\% & 11.23  & 0.17  & 3.35  & 28.42  & 0.647  \\
          &       & 0.5   & 11.75\% & 11.17\% & 77.08\% & 11.21  & 0.13  & 3.32  & 30.13  & 0.648  \\
          &       & 0.7   & 7.97\% & 8.79\% & 83.24\% & 8.56  & 0.10  & 2.52  & 30.70  & 0.641  \\
\cmidrule{2-11}          & \multirow{3}[2]{*}{0.3} & 0.3   & 3.84\% & 8.97\% & 87.20\% & 7.25  & 0.09  & 2.17  & 29.29  & 0.653  \\
          &       & 0.5   & 5.77\% & 5.27\% & 88.95\% & 7.86  & 0.09  & 2.40  & 29.30  & 0.640  \\
          &       & 0.7   & 2.49\% & 3.31\% & 94.20\% & 6.74  & 0.09  & 2.07  & 29.67  & 0.645  \\
\cmidrule{2-11}          & \multirow{3}[2]{*}{0.4} & 0.3   & 1.30\% & 3.53\% & 95.17\% & 6.45  & 0.07  & 1.95  & 29.49  & 0.645  \\
          &       & 0.5   & 0.94\% & 3.39\% & 95.67\% & 6.43  & 0.08  & 1.96  & 29.19  & 0.633  \\
          &       & 0.7   & 2.19\% & 2.18\% & 95.62\% & 6.64  & 0.08  & 2.08  & 29.36  & 0.638  \\
\cmidrule{2-11}          & \multirow{3}[2]{*}{0.5} & 0.3   & 0.94\% & 1.46\% & 97.60\% & 6.42  & 0.08  & 1.99  & 29.39  & 0.641  \\
          &       & 0.5   & 0.25\% & 0.80\% & 98.95\% & 6.40  & 0.08  & 2.03  & 29.30  & 0.645  \\
          &       & 0.7   & 0.08\% & 0.84\% & 99.07\% & 6.40  & 0.07  & 1.96  & 29.58  & 0.655  \\
    \midrule
    \multirow{12}[8]{*}{0.5} & \multirow{3}[2]{*}{0.2} & 0.3   & 13.33\% & 7.31\% & 79.37\% & 12.08  & 0.14  & 3.57  & 30.33  & 0.536  \\
          &       & 0.5   & 14.87\% & 8.61\% & 76.52\% & 10.17  & 0.11  & 2.98  & 31.54  & 0.507  \\
          &       & 0.7   & 8.40\% & 5.06\% & 86.54\% & 8.90  & 0.11  & 2.72  & 31.05  & 0.587  \\
\cmidrule{2-11}          & \multirow{3}[2]{*}{0.3} & 0.3   & 5.42\% & 4.28\% & 90.31\% & 7.73  & 0.09  & 2.29  & 30.17  & 0.533  \\
          &       & 0.5   & 4.59\% & 2.31\% & 93.10\% & 7.34  & 0.09  & 2.24  & 30.16  & 0.510  \\
          &       & 0.7   & 4.80\% & 2.79\% & 92.40\% & 7.37  & 0.09  & 2.20  & 29.85  & 0.542  \\
\cmidrule{2-11}          & \multirow{3}[2]{*}{0.4} & 0.3   & 2.12\% & 1.66\% & 96.22\% & 6.73  & 0.08  & 2.02  & 29.93  & 0.578  \\
          &       & 0.5   & 3.48\% & 2.61\% & 93.91\% & 6.53  & 0.08  & 2.00  & 30.15  & 0.490  \\
          &       & 0.7   & 2.02\% & 1.14\% & 96.84\% & 6.72  & 0.08  & 2.01  & 29.97  & 0.543  \\
\cmidrule{2-11}          & \multirow{3}[2]{*}{0.5} & 0.3   & 1.41\% & 0.90\% & 97.69\% & 6.40  & 0.08  & 2.02  & 29.97  & 0.560  \\
          &       & 0.5   & 1.24\% & 1.29\% & 97.47\% & 6.38  & 0.07  & 1.96  & 29.75  & 0.573  \\
          &       & 0.7   & 1.29\% & 0.99\% & 97.72\% & 6.43  & 0.08  & 2.05  & 30.45  & 0.571  \\
    \midrule
    \multirow{12}[8]{*}{1} & \multirow{3}[2]{*}{0.2} & 0.3   & 32.06\% & 19.56\% & 48.38\% & 23.53  & 0.27  & 6.77  & 29.74  & 0.731  \\
          &       & 0.5   & 23.13\% & 15.80\% & 61.06\% & 18.40  & 0.22  & 5.25  & 30.75  & 0.752  \\
          &       & 0.7   & 17.58\% & 15.04\% & 67.38\% & 11.72  & 0.14  & 3.35  & 29.76  & 0.740  \\
\cmidrule{2-11}          & \multirow{3}[2]{*}{0.3} & 0.3   & 12.48\% & 14.71\% & 72.82\% & 9.40  & 0.11  & 2.75  & 29.19  & 0.743  \\
          &       & 0.5   & 11.20\% & 8.39\% & 80.40\% & 11.59  & 0.13  & 3.28  & 29.06  & 0.750  \\
          &       & 0.7   & 7.32\% & 7.07\% & 85.61\% & 8.69  & 0.10  & 2.66  & 29.36  & 0.742  \\
\cmidrule{2-11}          & \multirow{3}[2]{*}{0.4} & 0.3   & 6.24\% & 7.43\% & 86.34\% & 7.15  & 0.08  & 2.14  & 29.48  & 0.740  \\
          &       & 0.5   & 3.86\% & 6.24\% & 89.90\% & 6.59  & 0.08  & 2.00  & 29.13  & 0.742  \\
          &       & 0.7   & 5.82\% & 4.37\% & 89.81\% & 7.32  & 0.09  & 2.28  & 29.24  & 0.730  \\
\cmidrule{2-11}          & \multirow{3}[2]{*}{0.5} & 0.3   & 2.27\% & 2.59\% & 95.14\% & 6.41  & 0.08  & 1.99  & 29.14  & 0.719  \\
          &       & 0.5   & 1.69\% & 3.02\% & 95.30\% & 6.50  & 0.08  & 2.01  & 29.42  & 0.743  \\
          &       & 0.7   & 0.69\% & 2.99\% & 96.32\% & 6.40  & 0.07  & 1.97  & 29.64  & 0.740  \\
    \midrule
    \multirow{12}[8]{*}{2} & \multirow{3}[2]{*}{0.2} & 0.3   & 30.37\% & 17.40\% & 52.23\% & 22.84  & 0.32  & 6.77  & 29.87  & 0.756  \\
          &       & 0.5   & 27.08\% & 17.14\% & 55.79\% & 17.49  & 0.20  & 5.03  & 30.42  & 0.757  \\
          &       & 0.7   & 23.69\% & 17.63\% & 58.69\% & 13.59  & 0.16  & 3.93  & 29.31  & 0.755  \\
\cmidrule{2-11}          & \multirow{3}[2]{*}{0.3} & 0.3   & 14.34\% & 14.08\% & 71.59\% & 10.78  & 0.13  & 3.15  & 29.11  & 0.756  \\
          &       & 0.5   & 17.67\% & 13.65\% & 68.68\% & 12.16  & 0.14  & 3.61  & 29.28  & 0.757  \\
          &       & 0.7   & 11.54\% & 8.66\% & 79.79\% & 8.95  & 0.11  & 2.80  & 29.39  & 0.754  \\
\cmidrule{2-11}          & \multirow{3}[2]{*}{0.4} & 0.3   & 7.47\% & 9.05\% & 83.49\% & 7.97  & 0.09  & 2.38  & 29.56  & 0.744  \\
          &       & 0.5   & 4.63\% & 5.80\% & 89.57\% & 7.09  & 0.08  & 2.13  & 29.14  & 0.747  \\
          &       & 0.7   & 6.38\% & 5.98\% & 87.64\% & 7.83  & 0.09  & 2.39  & 29.48  & 0.754  \\
\cmidrule{2-11}          & \multirow{3}[2]{*}{0.5} & 0.3   & 2.17\% & 3.22\% & 94.61\% & 6.66  & 0.08  & 2.07  & 29.58  & 0.732  \\
          &       & 0.5   & 1.03\% & 2.38\% & 96.59\% & 6.57  & 0.08  & 1.97  & 29.30  & 0.757  \\
          &       & 0.7   & 0.95\% & 3.56\% & 95.49\% & 6.43  & 0.07  & 1.94  & 29.26  & 0.759  \\
    \bottomrule
    \end{tabular}%
  \caption{Experimental Results on Cifar10.}
  \label{tab:ref_cifar}%
\end{table}%
\begin{table}[htbp]
  \centering
    \begin{tabular}{ccccccccccc}
    \toprule
    \multicolumn{3}{c}{\textbf{Hyperparameters}} & \multicolumn{3}{c}{\textbf{Mask ratio}} & \multicolumn{4}{c}{\textbf{Time Cost (s)}} & \textbf{Performance} \\
    \midrule
    \midrule
    $\alpha$  & $\tau$  & $\rho$  &  $M_{\text{enc}}$ &  $M_{\text{pers}}$ &  $M_{\text{noise}}$ & Encryption & Decryption & Aggregation & Training & Acc \\
    \midrule
    \multirow{12}[8]{*}{0.2} & \multirow{3}[2]{*}{0.2} & 0.3   & 21.40\% & 14.52\% & 64.08\% & 19.23  & 0.22  & 5.50  & 30.13  & 0.341  \\
          &       & 0.5   & 25.66\% & 17.10\% & 57.24\% & 15.80  & 0.18  & 4.53  & 29.88  & 0.013  \\
          &       & 0.7   & 29.76\% & 18.66\% & 51.58\% & 19.00  & 0.22  & 5.43  & 30.90  & 0.396  \\
\cmidrule{2-11}          & \multirow{3}[2]{*}{0.3} & 0.3   & 27.43\% & 17.56\% & 55.01\% & 22.98  & 0.28  & 6.47  & 30.02  & 0.418  \\
          &       & 0.5   & 14.86\% & 10.72\% & 74.42\% & 12.22  & 0.14  & 3.75  & 30.89  & 0.342  \\
          &       & 0.7   & 5.55\% & 1.01\% & 93.44\% & 11.42  & 0.13  & 3.32  & 30.37  & 0.010  \\
\cmidrule{2-11}          & \multirow{3}[2]{*}{0.4} & 0.3   & 18.04\% & 11.68\% & 70.29\% & 13.88  & 0.16  & 3.97  & 30.27  & 0.393  \\
          &       & 0.5   & 19.22\% & 12.14\% & 68.63\% & 17.65  & 0.21  & 5.17  & 30.87  & 0.417  \\
          &       & 0.7   & 14.72\% & 12.27\% & 73.01\% & 10.16  & 0.12  & 3.06  & 30.53  & 0.309  \\
\cmidrule{2-11}          & \multirow{3}[2]{*}{0.5} & 0.3   & 13.03\% & 2.28\% & 84.69\% & 10.00  & 0.12  & 2.96  & 29.51  & 0.010  \\
          &       & 0.5   & 19.94\% & 16.04\% & 64.02\% & 10.61  & 0.12  & 3.12  & 29.11  & 0.409  \\
          &       & 0.7   & 21.12\% & 15.33\% & 63.55\% & 12.39  & 0.15  & 3.53  & 29.53  & 0.420  \\
    \midrule
    \multirow{12}[8]{*}{0.5} & \multirow{3}[2]{*}{0.2} & 0.3   & 14.34\% & 13.04\% & 72.63\% & 9.22  & 0.11  & 2.74  & 29.64  & 0.311  \\
          &       & 0.5   & 9.89\% & 1.92\% & 88.19\% & 9.49  & 0.11  & 2.80  & 29.12  & 0.308  \\
          &       & 0.7   & 10.42\% & 13.00\% & 76.58\% & 9.13  & 0.11  & 2.66  & 29.58  & 0.397  \\
\cmidrule{2-11}          & \multirow{3}[2]{*}{0.3} & 0.3   & 14.22\% & 12.56\% & 73.22\% & 9.90  & 0.12  & 2.93  & 29.24  & 0.414  \\
          &       & 0.5   & 10.36\% & 8.15\% & 81.49\% & 9.22  & 0.11  & 2.73  & 29.81  & 0.333  \\
          &       & 0.7   & 5.15\% & 1.50\% & 93.35\% & 8.51  & 0.10  & 2.54  & 29.19  & 0.011  \\
\cmidrule{2-11}          & \multirow{3}[2]{*}{0.4} & 0.3   & 14.64\% & 11.17\% & 74.19\% & 9.78  & 0.12  & 2.86  & 29.68  & 0.404  \\
          &       & 0.5   & 8.90\% & 6.51\% & 84.58\% & 10.47  & 0.12  & 3.19  & 29.81  & 0.416  \\
          &       & 0.7   & 5.96\% & 4.97\% & 89.06\% & 7.90  & 0.09  & 2.39  & 30.03  & 0.347  \\
\cmidrule{2-11}          & \multirow{3}[2]{*}{0.5} & 0.3   & 7.82\% & 1.67\% & 90.51\% & 7.71  & 0.09  & 2.33  & 29.48  & 0.359  \\
          &       & 0.5   & 9.47\% & 7.58\% & 82.95\% & 8.18  & 0.10  & 2.43  & 29.51  & 0.400  \\
          &       & 0.7   & 10.86\% & 8.13\% & 81.01\% & 8.02  & 0.09  & 2.39  & 29.58  & 0.415  \\
    \midrule
    \multirow{12}[8]{*}{1} & \multirow{3}[2]{*}{0.2} & 0.3   & 5.57\% & 4.43\% & 90.00\% & 7.21  & 0.08  & 2.16  & 29.84  & 0.350  \\
          &       & 0.5   & 8.55\% & 1.98\% & 89.48\% & 8.06  & 0.09  & 2.34  & 29.07  & 0.010  \\
          &       & 0.7   & 6.21\% & 6.04\% & 87.76\% & 7.00  & 0.09  & 2.18  & 29.46  & 0.394  \\
\cmidrule{2-11}          & \multirow{3}[2]{*}{0.3} & 0.3   & 9.45\% & 9.85\% & 80.70\% & 7.67  & 0.09  & 2.28  & 29.22  & 0.415  \\
          &       & 0.5   & 4.07\% & 6.70\% & 89.24\% & 6.74  & 0.08  & 2.08  & 30.12  & 0.337  \\
          &       & 0.7   & 6.69\% & 1.35\% & 91.96\% & 7.70  & 0.09  & 2.41  & 29.19  & 0.179  \\
\cmidrule{2-11}          & \multirow{3}[2]{*}{0.4} & 0.3   & 4.30\% & 5.96\% & 89.74\% & 6.73  & 0.08  & 2.01  & 29.49  & 0.393  \\
          &       & 0.5   & 5.39\% & 6.92\% & 87.69\% & 6.81  & 0.08  & 2.08  & 29.79  & 0.412  \\
          &       & 0.7   & 4.21\% & 3.77\% & 92.02\% & 7.08  & 0.08  & 2.15  & 29.54  & 0.330  \\
\cmidrule{2-11}          & \multirow{3}[2]{*}{0.5} & 0.3   & 0.80\% & 0.22\% & 98.97\% & 6.94  & 0.08  & 2.15  & 29.00  & 0.334  \\
          &       & 0.5   & 3.66\% & 3.23\% & 93.11\% & 7.15  & 0.08  & 2.23  & 29.61  & 0.412  \\
          &       & 0.7   & 4.14\% & 2.60\% & 93.26\% & 7.29  & 0.09  & 2.22  & 29.82  & 0.416  \\
    \midrule
    \multirow{12}[8]{*}{2} & \multirow{3}[2]{*}{0.2} & 0.3   & 1.91\% & 2.08\% & 96.01\% & 6.53  & 0.08  & 2.12  & 29.71  & 0.343  \\
          &       & 0.5   & 0.97\% & 0.25\% & 98.78\% & 6.56  & 0.08  & 2.03  & 29.55  & 0.316  \\
          &       & 0.7   & 2.79\% & 2.63\% & 94.58\% & 6.55  & 0.08  & 2.01  & 29.40  & 0.401  \\
\cmidrule{2-11}          & \multirow{3}[2]{*}{0.3} & 0.3   & 3.02\% & 3.11\% & 93.87\% & 6.62  & 0.08  & 2.00  & 29.57  & 0.415  \\
          &       & 0.5   & 1.54\% & 1.89\% & 96.56\% & 6.44  & 0.07  & 1.99  & 30.12  & 0.330  \\
          &       & 0.7   & 2.05\% & 2.71\% & 95.23\% & 6.67  & 0.08  & 2.09  & 29.18  & 0.010  \\
\cmidrule{2-11}          & \multirow{3}[2]{*}{0.4} & 0.3   & 1.38\% & 2.19\% & 96.43\% & 6.42  & 0.08  & 2.01  & 28.99  & 0.399  \\
          &       & 0.5   & 1.82\% & 2.14\% & 96.04\% & 6.50  & 0.08  & 2.01  & 29.10  & 0.411  \\
          &       & 0.7   & 0.75\% & 2.28\% & 96.98\% & 6.43  & 0.07  & 1.98  & 29.54  & 0.340  \\
\cmidrule{2-11}          & \multirow{3}[2]{*}{0.5} & 0.3   & 1.51\% & 0.65\% & 97.84\% & 6.76  & 0.08  & 2.13  & 29.26  & 0.371  \\
          &       & 0.5   & 1.25\% & 2.66\% & 96.10\% & 6.29  & 0.07  & 1.93  & 29.01  & 0.401  \\
          &       & 0.7   & 2.97\% & 2.70\% & 94.33\% & 6.38  & 0.07  & 1.96  & 29.27  & 0.417  \\
    \bottomrule
    \end{tabular}%
  \caption{Experimental Results on Cifar100.}
  \label{tab:ref_cifar100}%
\end{table}%
\begin{table}[htbp]
  \centering
    \begin{tabular}{ccccccccccc}
    \toprule
    \multicolumn{3}{c}{\textbf{Hyperparameters}} & \multicolumn{3}{c}{\textbf{Mask ratio}} & \multicolumn{4}{c}{\textbf{Time Cost (s)}} & \textbf{Performance} \\
    \midrule
    \midrule
    $\alpha$  & $\tau$  & $\rho$  &  $M_{\text{enc}}$ &  $M_{\text{pers}}$ &  $M_{\text{noise}}$ & Encryption & Decryption & Aggregation & Training & Acc \\
    \midrule
    \multirow{12}[6]{*}{0.2} & \multirow{4}[2]{*}{0.02} & 0.2   & 17.36\% & 12.10\% & 70.54\% & 13.55  & 0.15  & 3.89  & 51.65  & 0.722  \\
          &       & 0.3   & 10.97\% & 8.84\% & 80.19\% & 11.34  & 0.13  & 3.32  & 53.16  & 0.757  \\
          &       & 0.5   & 4.91\% & 7.49\% & 87.60\% & 7.26  & 0.08  & 2.19  & 52.20  & 0.770  \\
          &       & 0.7   & 2.69\% & 8.93\% & 88.38\% & 6.55  & 0.08  & 1.96  & 50.24  & 0.750  \\
\cmidrule{2-11}          & \multirow{4}[2]{*}{0.05} & 0.2   & 5.66\% & 4.75\% & 89.59\% & 8.30  & 0.10  & 2.50  & 50.93  & 0.753  \\
          &       & 0.3   & 3.90\% & 3.92\% & 92.17\% & 7.42  & 0.09  & 2.23  & 50.48  & 0.743  \\
          &       & 0.5   & 2.38\% & 4.06\% & 93.55\% & 6.49  & 0.08  & 1.98  & 50.41  & 0.765  \\
          &       & 0.7   & 0.61\% & 2.82\% & 96.57\% & 6.42  & 0.08  & 2.00  & 50.97  & 0.758  \\
\cmidrule{2-11}          & \multirow{4}[2]{*}{0.1} & 0.2   & 2.64\% & 2.21\% & 95.15\% & 6.94  & 0.08  & 2.09  & 50.61  & 0.758  \\
          &       & 0.3   & 1.62\% & 1.60\% & 96.79\% & 6.56  & 0.08  & 2.15  & 50.83  & 0.703  \\
          &       & 0.5   & 0.66\% & 1.41\% & 97.93\% & 6.41  & 0.07  & 1.94  & 50.43  & 0.758  \\
          &       & 0.7   & 0.20\% & 1.41\% & 98.39\% & 6.32  & 0.07  & 1.93  & 49.44  & 0.747  \\
    \midrule
    \multirow{12}[6]{*}{0.5} & \multirow{4}[2]{*}{0.02} & 0.2   & 23.01\% & 15.82\% & 61.17\% & 16.72  & 0.19  & 5.03  & 48.81  & 0.874  \\
          &       & 0.3   & 19.81\% & 13.48\% & 66.70\% & 14.35  & 0.17  & 4.17  & 49.25  & 0.877  \\
          &       & 0.5   & 8.72\% & 10.53\% & 80.74\% & 8.91  & 0.11  & 2.62  & 47.64  & 0.871  \\
          &       & 0.7   & 1.49\% & 14.71\% & 83.80\% & 6.61  & 0.08  & 2.00  & 49.07  & 0.862  \\
\cmidrule{2-11}          & \multirow{4}[2]{*}{0.05} & 0.2   & 9.51\% & 7.51\% & 82.97\% & 10.09  & 0.12  & 3.01  & 47.87  & 0.879  \\
          &       & 0.3   & 6.21\% & 5.67\% & 88.13\% & 7.67  & 0.09  & 2.36  & 47.28  & 0.865  \\
          &       & 0.5   & 1.81\% & 5.53\% & 92.66\% & 6.63  & 0.08  & 2.04  & 47.76  & 0.871  \\
          &       & 0.7   & 0.32\% & 4.48\% & 95.20\% & 6.46  & 0.07  & 1.96  & 47.45  & 0.881  \\
\cmidrule{2-11}          & \multirow{4}[2]{*}{0.1} & 0.2   & 3.17\% & 2.77\% & 94.06\% & 7.16  & 0.08  & 2.18  & 48.15  & 0.875  \\
          &       & 0.3   & 1.26\% & 1.77\% & 96.97\% & 6.40  & 0.07  & 1.97  & 47.38  & 0.874  \\
          &       & 0.5   & 0.78\% & 1.81\% & 97.41\% & 6.42  & 0.08  & 1.95  & 48.20  & 0.880  \\
          &       & 0.7   & 0.47\% & 1.57\% & 97.96\% & 6.17  & 0.07  & 1.88  & 47.30  & 0.872  \\
    \midrule
    \multirow{12}[6]{*}{0.1} & \multirow{4}[2]{*}{0.02} & 0.2   & 23.22\% & 15.75\% & 61.03\% & 19.75  & 0.22  & 5.52  & 49.34  & 0.857  \\
          &       & 0.3   & 12.99\% & 10.84\% & 76.17\% & 15.44  & 0.17  & 4.52  & 50.79  & 0.850  \\
          &       & 0.5   & 6.93\% & 10.52\% & 82.55\% & 9.01  & 0.10  & 2.65  & 48.36  & 0.862  \\
          &       & 0.7   & 3.02\% & 15.43\% & 81.55\% & 6.98  & 0.08  & 2.09  & 47.81  & 0.856  \\
\cmidrule{2-11}          & \multirow{4}[2]{*}{0.05} & 0.2   & 11.43\% & 8.89\% & 79.68\% & 10.88  & 0.13  & 3.15  & 49.09  & 0.854  \\
          &       & 0.3   & 6.95\% & 6.39\% & 86.66\% & 8.33  & 0.10  & 2.51  & 48.03  & 0.864  \\
          &       & 0.5   & 1.67\% & 4.21\% & 94.12\% & 6.56  & 0.08  & 2.07  & 48.64  & 0.862  \\
          &       & 0.7   & 0.77\% & 4.80\% & 94.43\% & 6.45  & 0.08  & 1.98  & 48.20  & 0.861  \\
\cmidrule{2-11}          & \multirow{4}[2]{*}{0.1} & 0.2   & 3.75\% & 3.41\% & 92.84\% & 7.69  & 0.09  & 2.35  & 48.00  & 0.850  \\
          &       & 0.3   & 2.00\% & 1.82\% & 96.19\% & 6.75  & 0.08  & 2.06  & 48.51  & 0.859  \\
          &       & 0.5   & 0.23\% & 1.31\% & 98.46\% & 6.40  & 0.07  & 1.93  & 48.70  & 0.857  \\
          &       & 0.7   & 0.18\% & 1.95\% & 97.87\% & 5.98  & 0.07  & 1.80  & 47.44  & 0.857  \\
    \bottomrule
    \end{tabular}%
  \caption{Experimental Results on SVHN.}
  \label{tab:ref_svhn}%
\end{table}%
\begin{table}[htbp]
  \centering
    \begin{tabular}{ccccccccccc}
    \toprule
    \multicolumn{3}{c}{\textbf{Hyperparameters}} & \multicolumn{3}{c}{\textbf{Mask ratio}} & \multicolumn{4}{c}{\textbf{Time Cost (s)}} & \textbf{Performance} \\
    \midrule
    \midrule
    $\alpha$  & $\tau$  & $\rho$  &  $M_{\text{enc}}$ &  $M_{\text{pers}}$ &  $M_{\text{noise}}$ & Encryption & Decryption & Aggregation & Training & Acc \\
    \midrule
    \multirow{12}[6]{*}{0.2} & \multirow{4}[2]{*}{0.02} & 0.2   & 3.37\% & 2.10\% & 94.53\% & 19.06  & 0.22  & 7.21  & 50.85  & 0.284  \\
          &       & 0.3   & 0.85\% & 0.47\% & 98.69\% & 10.68  & 0.13  & 4.90  & 51.73  & 0.290  \\
          &       & 0.5   & 1.82\% & 1.19\% & 97.00\% & 8.92  & 0.10  & 4.33  & 49.86  & 0.301  \\
          &       & 0.7   & 0.75\% & 0.81\% & 98.44\% & 8.39  & 0.10  & 4.12  & 50.14  & 0.316  \\
\cmidrule{2-11}          & \multirow{4}[2]{*}{0.05} & 0.2   & 1.16\% & 0.71\% & 98.12\% & 11.33  & 0.13  & 4.93  & 49.61  & 0.304  \\
          &       & 0.3   & 0.85\% & 0.47\% & 98.68\% & 8.23  & 0.10  & 4.55  & 50.22  & 0.289  \\
          &       & 0.5   & 0.64\% & 0.70\% & 98.65\% & 7.54  & 0.09  & 3.94  & 49.42  & 0.292  \\
          &       & 0.7   & 0.34\% & 0.84\% & 98.82\% & 7.23  & 0.09  & 3.89  & 51.23  & 0.290  \\
\cmidrule{2-11}          & \multirow{4}[2]{*}{0.1} & 0.2   & 0.46\% & 0.39\% & 99.15\% & 7.29  & 0.08  & 3.83  & 47.93  & 0.302  \\
          &       & 0.3   & 0.38\% & 0.28\% & 99.34\% & 6.46  & 0.07  & 3.61  & 47.32  & 0.293  \\
          &       & 0.5   & 0.20\% & 0.24\% & 99.56\% & 6.06  & 0.07  & 3.44  & 47.32  & 0.283  \\
          &       & 0.7   & 0.08\% & 0.25\% & 99.67\% & 6.08  & 0.07  & 3.44  & 46.91  & 0.293  \\
    \midrule
    \multirow{12}[6]{*}{0.5} & \multirow{4}[2]{*}{0.02} & 0.2   & 4.93\% & 3.29\% & 91.78\% & 22.25  & 0.26  & 7.97  & 50.54  & 0.429  \\
          &       & 0.3   & 1.64\% & 1.14\% & 97.22\% & 15.81  & 0.18  & 6.18  & 49.87  & 0.438  \\
          &       & 0.5   & 1.79\% & 2.21\% & 96.00\% & 9.46  & 0.11  & 4.40  & 48.88  & 0.444  \\
          &       & 0.7   & 0.42\% & 2.31\% & 97.27\% & 7.21  & 0.08  & 3.82  & 49.92  & 0.462  \\
\cmidrule{2-11}          & \multirow{4}[2]{*}{0.05} & 0.2   & 2.25\% & 1.76\% & 95.99\% & 11.69  & 0.13  & 5.01  & 49.39  & 0.425  \\
          &       & 0.3   & 1.50\% & 1.14\% & 97.36\% & 8.72  & 0.11  & 4.49  & 49.53  & 0.445  \\
          &       & 0.5   & 0.50\% & 1.25\% & 98.24\% & 7.21  & 0.09  & 3.91  & 49.22  & 0.443  \\
          &       & 0.7   & 0.17\% & 1.10\% & 98.73\% & 6.34  & 0.07  & 3.48  & 47.19  & 0.426  \\
\cmidrule{2-11}          & \multirow{4}[2]{*}{0.1} & 0.2   & 1.21\% & 1.03\% & 97.76\% & 7.50  & 0.09  & 3.82  & 46.19  & 0.452  \\
          &       & 0.3   & 0.44\% & 0.39\% & 99.16\% & 6.78  & 0.08  & 3.78  & 46.91  & 0.422  \\
          &       & 0.5   & 0.03\% & 0.48\% & 99.49\% & 6.00  & 0.07  & 3.42  & 46.61  & 0.423  \\
          &       & 0.7   & 0.12\% & 0.40\% & 99.48\% & 6.03  & 0.07  & 4.16  & 48.41  & 0.432  \\
    \midrule
    \multirow{12}[6]{*}{0.1} & \multirow{4}[2]{*}{0.02} & 0.2   & 7.25\% & 5.51\% & 87.25\% & 23.56  & 0.27  & 8.45  & 51.46  & 0.527  \\
          &       & 0.3   & 4.25\% & 3.23\% & 92.52\% & 14.38  & 0.17  & 5.96  & 49.94  & 0.510  \\
          &       & 0.5   & 1.91\% & 2.15\% & 95.94\% & 8.55  & 0.10  & 4.21  & 49.53  & 0.461  \\
          &       & 0.7   & 2.27\% & 1.87\% & 95.85\% & 7.80  & 0.09  & 4.24  & 49.72  & 0.538  \\
\cmidrule{2-11}          & \multirow{4}[2]{*}{0.05} & 0.2   & 1.67\% & 1.58\% & 96.75\% & 9.86  & 0.11  & 4.60  & 49.36  & 0.526  \\
          &       & 0.3   & 0.93\% & 0.67\% & 98.40\% & 8.30  & 0.10  & 4.27  & 49.82  & 0.525  \\
          &       & 0.5   & 0.65\% & 0.76\% & 98.58\% & 6.83  & 0.08  & 3.88  & 48.96  & 0.512  \\
          &       & 0.7   & 0.34\% & 0.70\% & 98.96\% & 6.68  & 0.08  & 3.84  & 51.03  & 0.527  \\
\cmidrule{2-11}          & \multirow{4}[2]{*}{0.1} & 0.2   & 0.58\% & 0.59\% & 98.83\% & 6.39  & 0.07  & 3.52  & 46.13  & 0.514  \\
          &       & 0.3   & 0.55\% & 0.43\% & 99.02\% & 6.30  & 0.07  & 3.52  & 46.60  & 0.501  \\
          &       & 0.5   & 0.23\% & 0.40\% & 99.37\% & 6.14  & 0.07  & 3.48  & 46.49  & 0.537  \\
          &       & 0.7   & 0.18\% & 0.40\% & 99.42\% & 6.55  & 0.08  & 3.85  & 49.99  & 0.504  \\
    \bottomrule
    \end{tabular}%
  \caption{Experimental Results on stl10.}
  \label{tab:ref_stl10}%
\end{table}%

\end{document}